\def\year{2018}\relax
\documentclass[letterpaper]{article} %DO NOT CHANGE THIS
\usepackage{aaai18}  %Required
\usepackage{times}  %Required
\usepackage{helvet}  %Required
\usepackage{courier}  %Required
\usepackage{url}  %Required
\usepackage{graphicx}  %Required
\usepackage{booktabs} % For formal tables
\usepackage{amsmath,amsthm}
\usepackage{subcaption}
\usepackage{multirow}
\usepackage[flushleft]{threeparttable}
\usepackage{amsfonts}
\usepackage{color}

\newcommand{\mathbold}[1]{\ensuremath{\boldsymbol{\mathbf{#1}}}}

\newcommand\norm[1]{\left\lVert#1\right\rVert}

\usepackage{array}
\newcolumntype{L}[1]{>{\raggedright\let\newline\\\arraybackslash\hspace{0pt}}m{#1}}
\newcolumntype{C}[1]{>{\centering\let\newline\\\arraybackslash\hspace{0pt}}m{#1}}
\newcolumntype{R}[1]{>{\raggedleft\let\newline\\\arraybackslash\hspace{0pt}}m{#1}}

\newcommand{\mbz}{\mathbold{z}}
\newcommand{\mbw}{\mathbold{w}}

\newcommand{\cL}{\mathcal{L}}

\theoremstyle{plain}
\newtheorem{thm1}{Theorem} % reset theorem numbering for each chapter

\newtheorem{thm3}{Theorem}

\theoremstyle{definition}
\newtheorem{defn}[thm1]{Definition} % definition numbers are dependent on theorem numbers
\newtheorem{prob}[thm3]{Problem}
\usepackage[belowskip=-3.5pt,aboveskip=10pt]{caption}

\begin{document}
\title{Neural Ideal Point Estimation Network}
\author{
	Kyungwoo Song, Wonsung Lee, \and Il-Chul Moon \\
	Korea Advanced Institute of Science and Technology \\
	291 Daehak-ro, Yuseong-gu \\
	Daejeon 34141, South Korea \\
	\{gtshs2,aporia,icmoon\}@kaist.ac.kr \\
}
\maketitle
\begin{abstract}
Understanding politics is challenging because the politics take the influence from everything. Even we limit ourselves to the political context in the legislative processes; we need a better understanding of latent factors, such as legislators, bills, their ideal points, and their relations. From the modeling perspective, this is difficult 1) because these observations lie in a high dimension that requires learning on low dimensional representations, and 2) because these observations require complex probabilistic modeling with latent variables to reflect the causalities. This paper presents a new model to reflect and understand this political setting, NIPEN, including factors mentioned above in the legislation. We propose two versions of NIPEN: one is a hybrid model of deep learning and probabilistic graphical model, and the other model is a neural tensor model. Our result indicates that NIPEN successfully learns the manifold of the legislative bill texts, and NIPEN utilizes the learned low-dimensional latent variables to increase the prediction performance of legislators' votings. Additionally, by virtue of being a domain-rich probabilistic model, NIPEN shows the hidden strength of the legislators' trust network and their various characteristics on casting votes.
\end{abstract}

\section{Introduction}
Recent developments in machine learning have enabled a deeper understanding of human behavior in diverse contexts. These advances include divulging intentions and sentiments in dialogs \cite{berteroreal}; predicting purchases from online markets \cite{chong2015predicting}; recommending movies to friends \cite{shah2017matrix}; and discovering social network links between individuals \cite{Guo2015}. The recent machine learning models provide the contexts of these behaviors, which have been regarded as the latent aspects of human behavior. 
%The parameter inference, or {\it learning}, estimates appropriate latent distributions, and this tells why a human behaves in a certain way.

One latent modeling of human behavior can be a form of complex Bayesian probabilistic models, a.k.a. probabilistic graphical model (PGM). The modelers used graphical notations, embedding the probabilistic variables and their causalities, to represent the key factors and their relations. For instance, latent Dirichlet allocation (LDA) models the generative process of documents, i.e. the composition of topics at large, a main topic of documents, and a word selection when describing a topic \cite{Blei2003}. 
%Now, the community has expanded models, such as correlating collaborative filtering and topics to better recommend movies or research papers \cite{ChongWang2011,Purushotham2012}; explicitly separating the topics and the sentiments for joint sentiment classification and topic clustering, etc.

Another effort in modeling the latent variable is improving the quality of the latent representation of the data. While the above probabilistic models focused on the contextual modeling, the latent variables reside in a high dimensional and nonlinear space, so the learning of the latent variables have been limited. 
%The recent development in the deep learning community has advanced in learning this manifold space with various autoencoders. 
For example, the stacked de-noising autoencoder (SDAE) \cite{VincentPASCALVINCENT2010} learns this manifold space through encoding the noised inputs into the low dimensional latent representations; and reconstructing the original inputs with the latent representations with neural network layers. Further advances have made through casting this autoencoding mechanism to the variational inference approaches, and a variational autoencoder (VAE) \cite{Kingma2013} optimizes the variational distribution of the latent representations with neural networks.

Supported by the two research advances, one distinct research direction has been merging the latent representation learning and the probabilistic graphical model on human behavior. Collaborative deep learning (CDL) \cite{Wang2015} is one example merging SDAE with a probabilistic model of matrix factorization that often used to explain and predict the human behavior of recommendations. Whereas CDL gives a clear passway on how we can further develop various models of human behavior with support from the deep learning, different application domains require different latent modeling, so the model structure needs to be further customized and expanded. 
%Additionally, the variants of autoencoders need to be tested in conjunction with the new model structures to discover which combination is better and why.

\begin{figure*}[t!]
	\centering
	\includegraphics[width=7in]{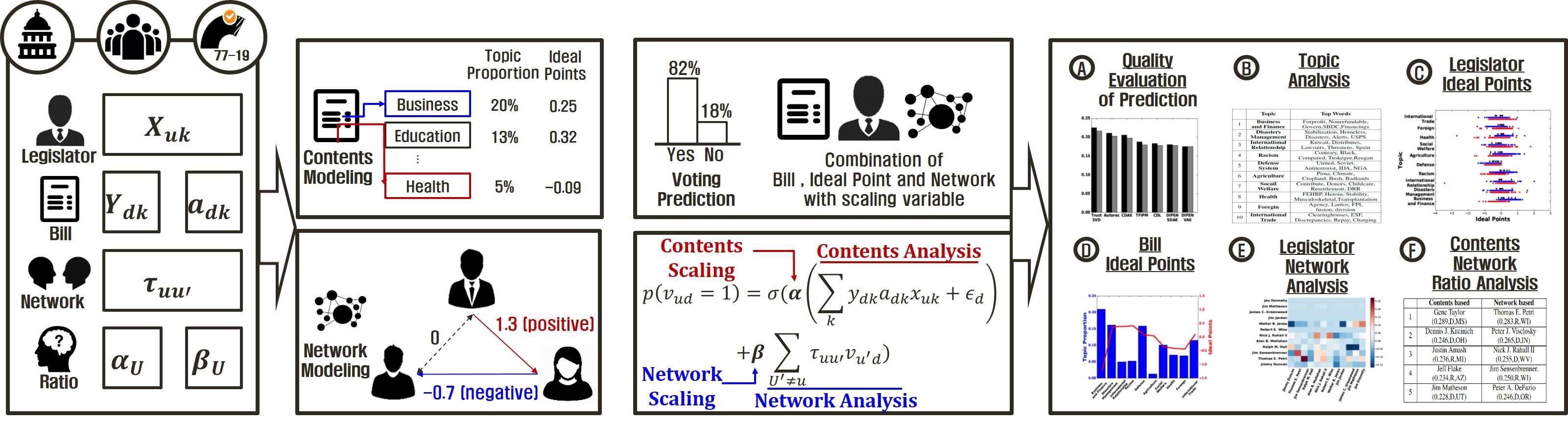}
	\caption{The summarized procedure of NIPEN. NIPEN predicts the votes with the combination of contents and network analyses. We can interpret not only an individual legislator's ideal points but also trust networks between legislators}
	\label{fig:flow_diagram}
\end{figure*}

This paper introduces Neural Ideal Point Estimation Network (NIPEN) which models the generative process of political voting by estimating ideal points in diverse legislative aspects with learning the low dimensional representations from neural networks. Specifically, we propose two versions of NIPEN. The first version, NIPEN-PGM is a hybrid model by representing the contextual causalities as a PGM, and by learning the low dimensional representations with multi-layered perceptron (MLP) autoencoders, i.e. SDAE and VAE. The second version, NIPEN-Tensor, is a neural tensor model that substitutes the PGM part with the neural tensor model. NIPEN-Tensor could be viewed as a generalized version of NIPEN-PGM. NIPEN-Tensor models the legislative voting with the tensor composition and the nonlinear operations between diverse legislative factors while NIPEN-PGM assumes the marginalization and the linearized operation in the same modeling part. 
%For text-rich datasets, NIPEN-Tensor demonstrates a better prediction and a fitting result because the tensor composition and the nonlinear operations provide a better modeling of complex context, which is critical in applying machine learning methods to diverse social domains.

Second, NIPEN is the most comprehensive model in the latent modeling of the political domain. Assuming that we model a voting process of legislators, NIPEN is the first model of unifying 1) the voting behavior, 2) the network influence between congressmen, 3) the political ideal point of bills and congressmen, 4) the textual topic of bills, and 5) the relative strength of network influence and ideal points when casting a vote. Some of these latent variables have been seen in other models, \cite{Gerrish2012,Gu2014,Chaney2015}, but not as the unified model to depict a whole political picture. Since diverse factors, such as the contents of the bill and the human relations, greatly influence the voting \cite{cohen2014friends}, an effective modeling of the legislative voting requires an integrated model, such as NIPEN. We show that NIPEN recorded significant performance improvements in all metrics compared to existing models. We also show various qualitative analyses that can only obtained via this comprehensive model. The entire procedures and analyses
of NIPEN is summarized by Figure \ref{fig:flow_diagram}.

%\cite{Gerrish2012,Gu2014,Nguyen2015,Chaney2015}

%The main contributions of NIPEN comparing with previous researches can be summarized as below.
%\begin{itemize}
%	\item NIPEN is the first model which incorporate VAE and MF with document modeling. We derive a unified EM-style learning algorithm for the parameter inference. This VAE integrated MF showed superior performances than MF with SDAE.
%	\item We propose various novel ideal point estimation models (NIPEN-SDAE, NIPEN-VAE, NIPEN-VAE (approx.)) which consider bills, legislator, and legislator's network simultaneously, which makes NIPEN as the first model to analyze the most comprehensive political picture. Furthermore, we also present a NIPEN-tensor model that generalizes existing ideal points analysis models.
%	\item Our proposed model, NIPEN outperforms other state-of-the-arts models on various measures. The experiments were conducted on two real-US-congress data. We collected and preprocessed a new roll-call dataset and opened it to the public.
%\end{itemize}

\section{Previous Research}
\subsection{Modeling Political Network and Ideal Points}
Network analyses and ideal point estimation have been widely studied in computer science and quantitative political science for its importance. In the line of political network analyses, most studies analyzed co-sponsorship data \cite{Faust2002,Fowler2006}. Faust and Skvoretz (2002) clarified the topological structures in the network of the U.S. Senate (1973-1974), and they found that the network among U.S. Senator in 93rd Congress is O-star, I-star and Trans structure \cite{Faust2002}. Fowler (2006) inferred the relationship in U.S. Congress (1973-2004) by measuring the centrality to find the most central legislators \cite{Fowler2006}. In the community of ideal point estimation, Poole and Rosenthal (1985) proposed a nonl-inear logit model to account for political choices of legislators \cite{Poole1985}. However, it was a one-dimensional estimation, and the analysis could not identify what the ideal dimension stands for. To overcome the limitation, Clinton et al. (2004) proposed a multi-dimensional ideal point estimation model, but these models still remained at the simple logit model extensions \cite{Heckman1996,Clinton2004}.

With the advance of topic modeling, multi-dimensional ideal point models were developed, and these models provide more accurate interpretations on the ideal points. Gerrish and Blei (2012) proposed an issue-adjusted model \cite{Gerrish2012} with the labeled LDA \cite{Ramage2009}, and Yupeng et al. (2014) proposed a topic-factorized ideal point model (TFIPM) \cite{Gu2014} with probabilistic latent semantic analysis (PLSA) \cite{Hofmann1999} to estimate the ideal points of legislators based on roll-call data. Further extensions of TFIPM have made through including available domain data. For instance, Islam et al. (2016) proposed SCIPM by including co-sponsorship networks between judges in the supreme court \cite{Islam2016}. These works have remained in the extension of the probabilistic graphical model without the innovation from the deep learning community, which our work extends 1) the probabilistic graphical model with variational autoencoders and 2) the neural tensor model for the causality modeling of the legislative voting.
% Nguyen expands the TFIPM by incorporating the legislators' speech contents \cite{Nguyen2015}. In addition,

\subsection{Collaborative Filtering and Deep Learning}
Collaborative Filtering is a recommendation algorithm that considers the relationship between users and items \cite{Koren2009}. One of representative approach is a matrix factorization which factorizes the rating matrix as user latent and item latent factors.Recently, the deep learning has initiated two theoretic developments. First, the matrix factorization itself is a low-dimensional representation method because of its latent vector learning, so does the autoencoding in the deep learning. For example, Sedhain et al. (2015) proposed Autorec \cite{Sedhain2015}, a basic autoencoder based CF algorithm, and Autorec outperforms other state-of-the-art MF algorithms like LLORMA \cite{Lee2013}. Wu et al. (2016) expand Autorec by concatenating a user latent variable to the rating input information in the encoder part of Autorec \cite{Wu2016}. Li et al. (2015) adopted two autoencoders corresponding to users and items \cite{Fu2015}, and they showed the interaction mechanism between the two autoencoders by using the marginalized SDAE \cite{Chen2012}. Second, the matrix factorization is related to the low-dimensional feature representation by adding the representation of the model as the distilled version of the side information. For instance, Wang et al. (2015) proposed a collaborative deep learning (CDL) which combines SDAE with MF \cite{Wang2015}. Furthermore, Ying (2016) proposed a model of collaborative deep ranking which combines ranking with algorithm and SDAE \cite{Ying2016}. Wang et al. (2017) proposed the relational deep learning with SDAE to link prediction between items \cite{Wang2017}. 

\section{Method}
This section introduce the detailed descriptions of NIPEN-PGM and NIPEN-Tensor in turn. Appendix A formulates the assumptions and the research questions, and Appendix C enumerates all symbols in this study. 
%We first introduce NIPEN-PGM, and we show the adaptation of SDAE and VAE to the PGM for the legislative process. Second, we describe NIPEN-Tensor with emphasis on the changes made in the voting process within NIPEN-PGM. 

\subsection{NIPEN with Probabilistic Graphical Model and Autoencoders}
\begin{figure}[t!]
	\centering
	\includegraphics[width=3.2in,height=2.5in]{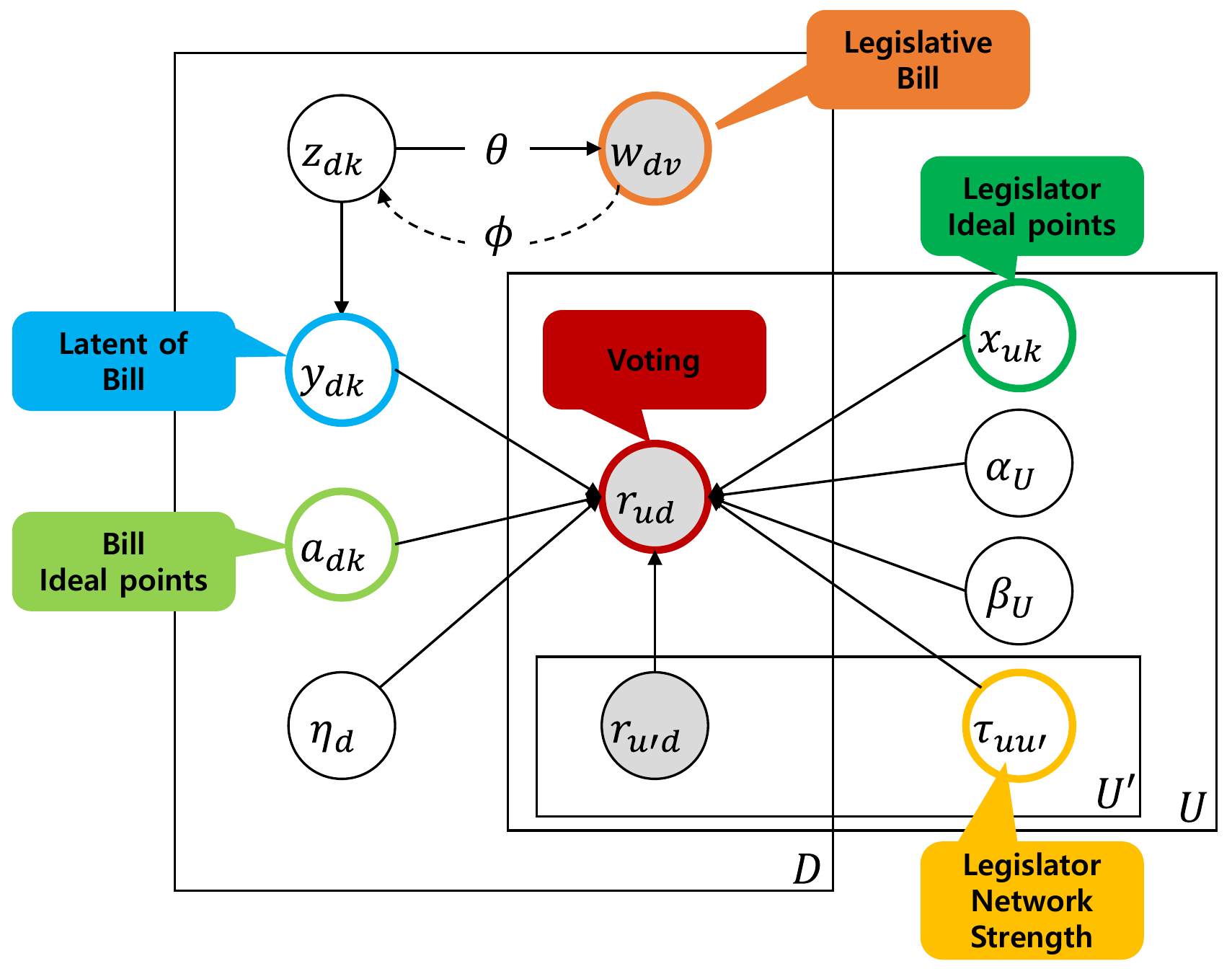}
	\caption{Graphical model representation of NIPEN-PGM}
	\label{fig:NIPEN_graphical_model}
\end{figure}

Figure \ref{fig:NIPEN_graphical_model} describes the model structure of NIPEN-PGM. We start the detailed description from the bill low dimension modeling part, which is the bill plate with the $d \in D$ subscript.
%NIPEN is a unified model of CF and an autoencoder. The unification is enabled by setting the item latent feature, $\mby$, by the addition of two components: the bill topic $\mbz$ and the latent offset $\xi$, which we follow the previous hybrid models, such as CTR and CDL. Therefore, we 
We apply either VAE or SDAE to learn the low dimensional representation, or topic, of $z_{dk}$\footnote{$d,u,$ and $k$ mean each document, legislator, topic respectively. Small subscripts indicate the row and column index in order.} with the observed bill text $w_{dv}$. $z_{dk}$ can be extracted through the probabilistic encoder, $q_{\phi}$ with parameter $\phi$ and decoder, $p_{\theta}$ with parameter $\theta$ which is further described in Appendix B. The topic representation of bills has two components: the bill latent $y_{dk}$ and the latent offset $\xi_{dk}$, and we model the combination of the two component as the below. 
\begin{equation*}  \begin{aligned} \label{eq:h_d}
y_{dk} = \xi_{dk} + z_{dk}, \quad \xi_{kd} \sim N(0,\lambda_y^{-1})
\end{aligned}  \end{equation*} 
Since the bill itself and the bill text may have two different latent variables, $\xi_{dk}$ becomes the \textit{offset} between the bill latent variable and the bill text latent variable, or \textit{topic}.

From the defined bill latent $y_{dk}$, we model how the bill latent generates the voting observation $r_{ud}$. Here, $u \in U$ is the dimension of the legislators. We assumed that a legislator cast votes considering three latent factors: the bill latent $y_{dk}$, the bill ideal point $a_{dk}$, and the legislators' ideal point $x_{uk}$. 
\begin{equation*}  \begin{aligned} \label{eq:ideal_point1}
a_{dk} \sim N(0,\lambda_u^{-1}), \quad x_{uk} \sim N(0,\lambda_u^{-1}) \\
\end{aligned}  \end{equation*}

Now, we define NIPEN-PGM without the network factor. This voting procedure is modeled as Eq. (\ref{eq:basic_ideal}) where $\eta_{d}$ is a bias value of a legislative bill, and $\sigma$ is a sigmoid function. Eq. (\ref{eq:basic_ideal}) is designed to increase the probability of voting \textit{YEA} when the ideal points of the bill and the legislator have the same sign; and when an ideal-aligned dimension of the bill latent variable is high. Additionally,  $\eta_{d}$ indicates whether the bill is more broadly accepted or not, regardless of ideal points. 
\begin{equation}  \begin{aligned} \label{eq:basic_ideal}
p(r_{ud}=1) =\sigma(\sum_{k=1}^{K}y_{dk}a_{dk}x_{uk}+\eta_{d})
\end{aligned}  \end{equation} 

Finally, we add the network component to NIPEN-PGM. The interest of a particular legislative group could be an important factor in the voting process. Following this implication, we modeled the network between two legislators as below. Before the network modeling, we limited the network influence between the legislators sharing the same term, and this neighbor set, $I_{u}$, is defined as a neighborhood of legislator, $u$.
\begin{equation*}  \begin{aligned} \label{eq:ideal_with_network}
\tau_{uu'} \sim N(0,\lambda_\tau^{-1}) \quad \alpha_{u} \sim N(0,\lambda_\alpha^{-1}) \quad \beta_{u} \sim N(0,\lambda_\alpha^{-1})
\end{aligned}  \end{equation*}
The legislator $u$'s voting is affected by two terms. The first term is the ideal alignment modeled in Eq. (\ref{eq:basic_ideal}). The second term is the voting record of the neighbor legislator, $r_{u'd}$, and the second term is also weighted by the network strength, $\tau_{uu'}$, between the two legislators. Since this is a linear summation, $\tau_{uu'}$ will model the degree of voting agreement between two legislators. These two terms are unified with scaling parameters $\alpha_{u}$ and $\beta_{u}$. The purpose of modeling $\alpha_{u}$ and $\beta_{u}$ is analyzing whether a certain legislator is influenced more either from the bill or from the network in casting votes. 

Eq. \ref{eq:ideal_point2} is the overall voting formulation of NIPEN-PGM.
\begin{equation}  \begin{aligned} \label{eq:ideal_point2}
\begin{split}
p(r_{ud}=1) &= \sigma(\alpha_{u}(\sum_{k}y_{dk}a_{dk}x_{uk}+\eta_{d}) \\
&+ \beta_{u}(\sum_{u'\in I_{u}}\tau_{uu'}r_{u'd})) \\
\end{split}
\end{aligned}  \end{equation}

\subsection{NIPEN with Neural Tensor Model}
Existing models, including NIPEN-PGM, do not directly model the relationships between the topics, which means that there is no cross-operiation between the dimension of $K$. Some cases, i.e. correlated topic model \cite{lafferty2006correlated}, model the correlation between topics via the logistic normal distribution, but this is not an operation modeling of topic influences, rather the variable modeling of topic covariance. 

\begin{figure}[t!]
	\centering
	\includegraphics[width=2.7in]{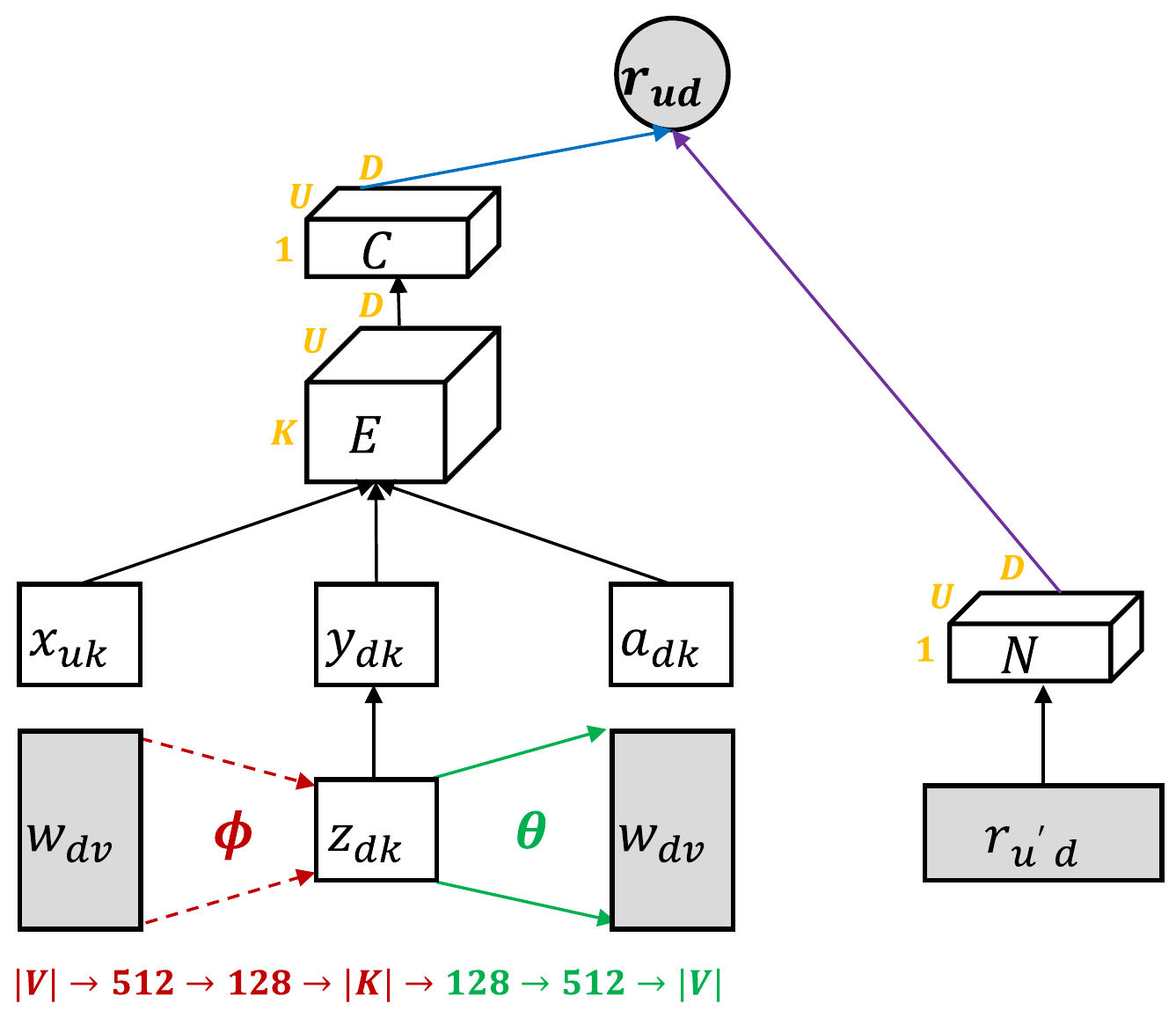}
	\caption{Neural network view of NIPEN-Tensor. The contents part is connected
		with the blue line (with content scaling parameter $\alpha_{u}$ ), and the network
		part is connected with the purple line (with the network scaling
		parameter $\beta_{u}$ ).}
	\label{fig:NIPEN_Tensor}
\end{figure}

The recent introduction of neural tensor models \cite{socher2013reasoning} enable the cross-operations between the latent topic dimension. This topic cross-operation can model the legislator's ideal point non-linear influences when two topics are combined within a bill. Here, we propose NIPEN-Tensor to incorporate the cross-topic influence in casting a vote, which could not be modeled in NIPEN-PGM. NIPEN-Tensor and NIPEN-PGM are similar in the parts of document and influence network modeling. 
The only different part is the voting decision modeled as Eq. \ref{eq:ideal_point2} which multiplies the factors per a topic and marginalizes. NIPEN-Tensor considers that the multiplication per a topic should be changed to consider the nonlinear effect from the topic set, not a single topic. Therefore, we represent the previous topic-wise multiplcaiton of $y_{dk}a_{dk}x_{uk}$ as a tensor $E$, and this tensor still treats the topic dimension to be independent. Then, we apply a fully-connected layer to cross-operate the topic dimension of $E$, and the neural network has $C$ that is the output of the cross-operation. The overall structure and formulation for the NIPEN-Tensor are shown in Figure \ref{fig:NIPEN_Tensor} and Eq. \ref{eq:NIPEN_Tensor}, respectively.
\begin{equation}  \begin{aligned} \label{eq:NIPEN_Tensor}
\begin{split}
E_{udk} &= x_{uk}y_{dk}z_{dk}\\
\widetilde{E}_{udl} &= \tanh(\sum_{k}E_{udk}W^{(T_{1})}_{kl} + b^{(T_{1})}_{l})\\
C_{ud} &=  \sum_{k}\widetilde{E}_{udl}W^{(T_{2})}_{l1} + \eta_{d} \\
N_{ud} &= \sum_{u'\in U}\tau_{uu'}v_{u'd} \\
p(r_{ud}=1) &= \sigma(\alpha_{u}C_{ud} + \beta_{u}\sum_{u'\in I_{u}}N_{u'd})
\end{split}
\end{aligned}
\end{equation}
$W^{(T_{1})}, b^{(T_{1})}, W^{(T_{2})}$ are weights and biases applied to $E_{udk}$, $\widetilde{E}_{udl}$ tensor. In particular, $W^{(T_{1})} \in R^{K \times K}$ models the correlation between topics, and $W^{(T_{2})} \in R^{K \times 1}$ models the influence of each topic on the voting. Since the signs of $x_{uk}, y_{dk}$, and $a_{dk}$ are important, we use $\tanh$ instead of ReLU (Rectified linear unit) to transform the outputs nonlinearly.

\subsection{Parameter Inference of NIPEN}
The parameters of both NIPENs are enumerated in the previous section, and we learn the parameters in two folds: learning the autoencoder to represent the bill topic and the CF, alternatively. The first set of parameters related to autoencoders is $\psi^{(1)}=(\theta,\phi)$; and the second set of parameters related with the legislative-CF is $\psi^{(2)}=(y,a,\eta,x,W^{(T_{1})},W^{(T_{2})},b^{(T_{1})},\tau,\alpha,\beta)$. 

The overall inference algorithm of both NIPENs follows the maximization of variational evidence lower bound with two assumptions. Following CDL, the first assumption is connecting the autoencoder and CF through $\xi$, and the strength is controlled by the variance of $\xi$, which is $\lambda_{y}$. When learning $\psi^{(1)}$, we apply the stochastic gradient variational Bayes (SGVB) estimator.   

Second, we assumed that the variational distribution of $\psi^{(2)}$ as a point mass for simplicity, so the parameters of the variational distribution are updated by each casted vote record, which is traditional Bayesian belief updates. Specifically, the likelihood of the posterior is presented as the lower bound in the below. Then, the lower bound, which has realized values of $q_{\phi}(\mbz|\mbw)$, $p_{\theta}(\mbz)$ and an observed input, has only $\psi^{(2)}$, so the gradient method can find the \textit{maximum a-posteriori}, or \textit{MAP}, of $\psi^{(2)}$.

As a summary, the objective function of both NIPENs is specified as follows:
\begin{equation*}  \begin{aligned} \label{eq:total_objective}
\begin{split}
\cL_{NIPEN} &= -{\operatorname{D_{KL}}}(q_{\phi} (\mbz | \mbw)\|p_{\theta}(\mbz)) + \frac{1}{L}\sum_{l=1}^{L} \log p_{\theta}(\mbw | \mbz^{l}) \\
&+\frac{\lambda_f}{2}\sum_{(u,d),r_{ud}\neq0}\frac{1+r_{ud}}{2}\log p(r_{ud}=1) \\
&+\frac{\lambda_f}{2}\sum_{(u,d),r_{ud}\neq0}\frac{1-r_{ud}}{2}\log p(r_{ud}=-1) \\
&-\frac{\lambda_y}{2} \sum_{d=1}^{D}\norm{y_{d} - z_{d}}^{2}_{2}
-\frac{\lambda_u}{2}(\norm{a}_{F}^{2} + \norm{x}_{F}^{2}) \\
&-\frac{\lambda_\tau}{2}(\norm{\tau}_{F}^{2}) -\frac{\lambda_\alpha}{2}(\norm{\alpha}_{2}^{2} + \norm{\beta}_{2}^{2})
\end{split}
\end{aligned} \end{equation*}
Similar to \cite{ChongWang2011,Wang2015}, the parameters related with the autoencoder and the legislative-CF are infered by coordinate ascents which maximizes $\cL_{NIPEN}$. For legislative-CF related parameters $\psi^{(2)}$, we take the gradient of $\cL_{NIPEN}$ w.r.t each parameters given the current $\theta$ and $\phi$. Given the legislative-CF related parameters $\psi^{(2)}$, we infer the autoencoder related parameters by computing $\nabla_{\psi^{(1)}}{\cL_{NIPEN}}$. We utilized the Tensorflow library \cite{abadi2016tensorflow} to optimize the parameters.

NIPEN-PGM and NIPEN-Tensor are only different in the vote casting process, and the related term in the objective function is the third and the fourth terms with $\log p(r_{ud}=1)$. These terms could be computed as the conventional gradient descent in two variants of NIPEN, so there is no change in the learning mechanism. 

In the original definition, the network, $\tau$, is a $|U|$-by-$|U|$ matrix, and the number of parameters becomes large given $O(U^{2})$. To reduce the squared complexity, $\tau$ is approximated by the product of $\widetilde{\tau_{1}}$ and $\widetilde{\tau_{2}}$ where $\widetilde{\tau_{1}}$ $\in  \mathbb{R}^{U \times G}$, $\widetilde{\tau_{2}}$ $\in  \mathbb{R}^{G \times U}$. We assume that $\widetilde{\tau_{1}}$ and $\widetilde{\tau_{2}}$ are not related. $G$ can be interpreted as the number of groups containing the legislators. This approximation results in $O(GU)$ for the network parameter inference.

\begin{table}[h!]
	\centering
	\caption{Attributes of \textit{Politic2013} and \textit{Politic2016} dataset}
	\label{table:dataset}
	\begin{tabular}{|C{3.8cm}|C{1.65cm}|C{1.65cm}|}
		\hline
		& \textit{Politic2013} & \textit{Politic2016} \\ \hline
		\# of legislators $(|U|)$                     &  1,540  &  1,537  \\ \hline
		\# of bills $(|D|)$                     &  7,162  &  7,975  \\ \hline
		\# of votings $(|D|)$                     &  2,779,703  &  2,999,844  \\ \hline
		\# of House                           &  1,299  &  1,266 \\ \hline
		\# of Senator                         &  241   &  271 \\ \hline
		\# of Republican                      &  767  &  778 \\ \hline
		\# of Democrat                        &  767  &  752 \\ \hline
		\# of unique word $(|V|)$                     &  10,000   & 13,581  \\ \hline
		{\begin{tabular}[c]{@{}c@{}}Average \# of unique word\\ for each bill
				$(\frac{\sum_{d,v}(I_{w_{dv}>0})}{V})$ \end{tabular}} & 192.77 & 378.66 \\ \hline
		{\begin{tabular}[c]{@{}c@{}}\# of bills less than \\ 10 unique words \end{tabular}} & 65 & 0 \\ \hline
		Period     &  1990-2013 & 1989-2016 \\ \hline
		Source     &  THOMAS   & GovTrack \\ \hline
		Data type & \multicolumn{2}{C{3.3cm}|}{1 (YEA),  -1 (NAY)} \\
		\hline
	\end{tabular}
\end{table}

\vspace{-\baselineskip}
\section{Results}
\subsection{Datasets on Political Ideal Points}
\begin{table*}[]
	\centering
	\caption{Quantitative evaluation on \textit{Politic2013} and \textit{Politic2016} datasets. Two-standard deviation is shown in parentheses}
	\label{table:quant1}
	\begin{tabular}{|C{2cm}|C{1.5cm}|C{1.5cm}|C{1.5cm}|C{1.5cm}|C{1.5cm}|C{1.5cm}|C{1.5cm}|C{1.5cm}|}
		\hline
		& \multicolumn{4}{c|}{\textit{Politic2013}}                                                                                                                                                                                                                       & \multicolumn{4}{c|}{\textit{Politic2016}}                                                                                                                                                                                                                       \\ \hline
		& RMSE     & MAE     &Accuracy  &NALL &RMSE      & MAE   & Accuracy  &NALL \\ \hline
		Trust SVD    & \begin{tabular}[c]{@{}c@{}}0.2253\\ ($\pm$0.0007)\end{tabular} & \begin{tabular}[c]{@{}c@{}}0.1399\\ ($\pm$0.0011)\end{tabular} & \begin{tabular}[c]{@{}c@{}}0.9408\\ ($\pm$0.0003)\end{tabular} & \begin{tabular}[c]{@{}c@{}}0.1866\\ ($\pm$0.0011)\end{tabular}      & \begin{tabular}[c]{@{}c@{}}0.2168\\ ($\pm$0.0011)\end{tabular} & \begin{tabular}[c]{@{}c@{}}0.1353\\ ($\pm$0.0010)\end{tabular} & \begin{tabular}[c]{@{}c@{}}0.9463\\ ($\pm$0.0009)\end{tabular} & \begin{tabular}[c]{@{}c@{}}0.1782\\ ($\pm$0.0015)\end{tabular}      \\ \hline
		Autorec             & \begin{tabular}[c]{@{}c@{}}0.2110\\ ($\pm$0.0099)\end{tabular} & \begin{tabular}[c]{@{}c@{}}0.0975\\ ($\pm$0.0136)\end{tabular} & \begin{tabular}[c]{@{}c@{}}0.9411\\ ($\pm$0.0056)\end{tabular} & \begin{tabular}[c]{@{}c@{}}0.1466\\ ($\pm$0.0177)\end{tabular}      & \begin{tabular}[c]{@{}c@{}}0.2031\\ ($\pm$0.0015)\end{tabular} & \begin{tabular}[c]{@{}c@{}}0.0886\\ ($\pm$0.0110)\end{tabular} & \begin{tabular}[c]{@{}c@{}}0.9454\\ ($\pm$0.0007)\end{tabular} & \begin{tabular}[c]{@{}c@{}}0.1349\\ ($\pm$0.0125)\end{tabular}      \\ \hline
		CDAE                & \begin{tabular}[c]{@{}c@{}}0.2059\\ ($\pm$0.0007)\end{tabular} & \begin{tabular}[c]{@{}c@{}}0.0831\\ ($\pm$0.0009)\end{tabular}  & \begin{tabular}[c]{@{}c@{}}0.9428\\ ($\pm$0.0006)\end{tabular}  & \begin{tabular}[c]{@{}c@{}}0.1450\\ ($\pm$0.0009)\end{tabular}       & \begin{tabular}[c]{@{}c@{}}0.1977\\ ($\pm$0.0037)\end{tabular}  & \begin{tabular}[c]{@{}c@{}}0.0802\\ ($\pm$0.0052)\end{tabular}  & \begin{tabular}[c]{@{}c@{}}0.9475\\ ($\pm$0.0023)\end{tabular}  & \begin{tabular}[c]{@{}c@{}}0.1357\\ ($\pm$0.0046)\end{tabular}       \\ \hline
		TFIPM               & \begin{tabular}[c]{@{}c@{}}0.1872\\ ($\pm$0.0002)\end{tabular}  & \begin{tabular}[c]{@{}c@{}}$0.0682^{\dagger}$\\ ($\pm$0.0002)\end{tabular}  & \begin{tabular}[c]{@{}c@{}}0.9526\\ ($\pm$0.0003)\end{tabular}  & \begin{tabular}[c]{@{}c@{}}0.1213\\ ($\pm$0.0007)\end{tabular}       & \begin{tabular}[c]{@{}c@{}}0.1794\\ ($\pm$0.0010)\end{tabular}  & \begin{tabular}[c]{@{}c@{}}$0.0625^{\dagger}$\\ ($\pm$0.0006)\end{tabular}  & \begin{tabular}[c]{@{}c@{}}0.9566\\ ($\pm$0.0005)\end{tabular}  & \begin{tabular}[c]{@{}c@{}}0.1121\\ ($\pm$0.0016)\end{tabular}       \\ \hline
		CDL                 & \begin{tabular}[c]{@{}c@{}}$0.1834^{\dagger}$\\ ($\pm$0.0008)\end{tabular}  & \begin{tabular}[c]{@{}c@{}}0.0786\\ ($\pm$0.0019)\end{tabular}  & \begin{tabular}[c]{@{}c@{}}$0.9554^{\dagger}$\\ ($\pm$0.0004)\end{tabular}  & \begin{tabular}[c]{@{}c@{}}$0.1147^{\dagger}$\\ ($\pm$0.0018)\end{tabular}       & \begin{tabular}[c]{@{}c@{}}$0.1780^{\dagger}$\\ ($\pm$0.0013)\end{tabular}  & \begin{tabular}[c]{@{}c@{}}0.0769\\ ($\pm$0.0012)\end{tabular}  & \begin{tabular}[c]{@{}c@{}}$0.9583^{\dagger}$\\ ($\pm$0.0008)\end{tabular}  & \begin{tabular}[c]{@{}c@{}}$0.1106^{\dagger}$\\ ($\pm$0.0017)\end{tabular}       \\ \hline

		\hline

		NIPEN-PGM(SDAE)          & \begin{tabular}[c]{@{}c@{}}0.1801**\\ ($\pm$0.0014)\end{tabular}  & \begin{tabular}[c]{@{}c@{}}0.0591**\\ ($\pm$0.0012)\end{tabular}  & \begin{tabular}[c]{@{}c@{}}0.9566**\\ ($\pm$0.0006)\end{tabular}  & \begin{tabular}[c]{@{}c@{}}0.1155\\ ($\pm$0.0018)\end{tabular}       & \begin{tabular}[c]{@{}c@{}}0.1779\\ ($\pm$0.0005)\end{tabular}  & \begin{tabular}[c]{@{}c@{}}\textbf{0.0560**}\\ ($\pm$0.0004)\end{tabular}  & \begin{tabular}[c]{@{}c@{}}0.9581\\ ($\pm$0.0003)\end{tabular}  & \begin{tabular}[c]{@{}c@{}}0.1173\\ ($\pm$0.0015)\end{tabular}       \\ \hline
		
		NIPEN-PGM(VAE, approx.) & \begin{tabular}[c]{@{}c@{}}0.1804\\ ($\pm$0.0089)\end{tabular}  & \begin{tabular}[c]{@{}c@{}}0.0611*\\ ($\pm$0.0065)\end{tabular}  & \begin{tabular}[c]{@{}c@{}}0.9565\\ ($\pm$0.0047)\end{tabular}  & \begin{tabular}[c]{@{}c@{}}0.1165\\ ($\pm$0.0086)\end{tabular}       & \begin{tabular}[c]{@{}c@{}}0.1791\\ ($\pm$0.0076)\end{tabular}  & \begin{tabular}[c]{@{}c@{}}0.0599\\ ($\pm$0.0057)\end{tabular}  & \begin{tabular}[c]{@{}c@{}}0.9571\\ ($\pm$0.0039)\end{tabular}  & \begin{tabular}[c]{@{}c@{}}0.1152\\ ($\pm$0.0070)\end{tabular}       \\ \hline
		
		NIPEN-PGM(VAE)           & \begin{tabular}[c]{@{}c@{}}\textbf{0.1753**}\\ ($\pm$0.0007)\end{tabular}  & \begin{tabular}[c]{@{}c@{}}\textbf{0.0588**}\\ ($\pm$0.0008)\end{tabular}  & \begin{tabular}[c]{@{}c@{}}\textbf{0.9587**}\\ ($\pm$0.0006)\end{tabular}  & \begin{tabular}[c]{@{}c@{}}\textbf{0.1075**}\\ ($\pm$0.0011)\end{tabular}       & \begin{tabular}[c]{@{}c@{}}0.1753**\\ ($\pm$0.0017)\end{tabular}  & \begin{tabular}[c]{@{}c@{}}0.0570**\\ ($\pm$0.0012)\end{tabular}  & \begin{tabular}[c]{@{}c@{}}0.9590**\\ ($\pm$0.0010)\end{tabular}  & \begin{tabular}[c]{@{}c@{}}0.1112\\ ($\pm$0.0024)\end{tabular}       \\ \hline
		
		NIPEN-Tensor           & \begin{tabular}[c]{@{}c@{}}0.1818**\\ ($\pm$0.0008)\end{tabular}  & \begin{tabular}[c]{@{}c@{}}0.0663**\\ ($\pm$0.0003)\end{tabular}  & \begin{tabular}[c]{@{}c@{}}0.9556**\\ ($\pm$0.0003)\end{tabular}  & \begin{tabular}[c]{@{}c@{}}0.1155\\ ($\pm$0.0020)\end{tabular}       & \begin{tabular}[c]{@{}c@{}}\textbf{0.1729**}\\ ($\pm$0.0015)\end{tabular}  & \begin{tabular}[c]{@{}c@{}}0.0608**\\ ($\pm$0.0006)\end{tabular}  & \begin{tabular}[c]{@{}c@{}}\textbf{0.9600**}\\ ($\pm$0.0008)\end{tabular}  & \begin{tabular}[c]{@{}c@{}}\textbf{0.1057**}\\ ($\pm$0.0022)\end{tabular}       \\ \hline

		\hline

		Improvement & 4.41\% & 13.78\% & 0.35\% & 6.27\% & 2.87\% & 10.40\% & 0.18\% & 4.43\% \\ \hline	
		
	\end{tabular}
	\begin{tablenotes}
	\small
	\item NALL : Negative Average Log Likelihood
	\item Improvement : Relative improvement of the best version of NIPEN compared to the best model, which is marked by $\dagger$, among the baselines
	\item $P^{*}<0.05; P^{**} < 0.01$ (Student's one-tailed $t$-test against the $\dagger$ model)
	\end{tablenotes}
\end{table*}

%\begin{figure*}
%	\centering
%	\subcaptionbox{RMSE\label{fig3:a}}{\includegraphics[width=1.7in]{RMSE.pdf}}\hspace{0em}%
%	\subcaptionbox{MAE\label{fig3:b}}{\includegraphics[width=1.7in]{MAE.pdf}}
%	%	\smallskip
%	\hspace{0em}%
%	\subcaptionbox{Accuracy\label{fig3:c}}{\includegraphics[width=1.7in]{Accuracy.pdf}}
%	\hspace{0em}%
%	\subcaptionbox{NALL \label{fig3:d}}{\includegraphics[width=1.7in]{Negative_Average_Loglikelihood.pdf}}
%	\hspace{0em}%
%	\caption{Quantitative comparison between baseline models and three types of NIPEN. Overall, NIPEN showed the best performance on all measures\label{fig:total_quant_compare}}
%\end{figure*}

We used two roll-call datasets, whose source is explained in Appendix D. Table \ref{table:dataset} provides the descriptive statistics of the two datasets: \textit{Politic2013} and \textit{Politic2016}. \textit{Politic2013} limits the number of a unique word to 10,000, and there are 65 bills which have less than ten words, while \textit{Politic2016} chooses 13,581 unique words, and there are no bills with less than ten words. \textit{Politic2013} is a more sparse dataset than \textit{Politic2016} in the ratings and the vocabulary sizes.

\subsection{Baselines and Implementation Details}
The variations of NIPEN were compared to five baseline models as follows:
\begin{itemize}
	
	\item \textbf{TFIPM}: Topic Factorized Ideal Point estimation Model \cite{Gu2014} is specialized in politics to analyze the roll-call data. 
%The model consists of PLSA \cite{Hofmann1999} and CF. The TFIPM links the PLSA part that models the document and the CF part with the weighted sum using the hyper-parameter lambda.
	
	\item \textbf{Autorec}: A simple autoencoder model which is utilized to predict the ratings. Autorec \cite{Sedhain2015} encodes and reconstructs the rating matrix. We used Item-based Autorec.
%, which has a better performance among User-based Autorec and Item-based Autorec.
%This model does not take side information in the prediction. 
	
	\item \textbf{Trust SVD}: Trust SVD \cite{Guo2015}, a type of trust-based matrix factorizations, is built on SVD++ with trust information. 
%Trust SVD assumes that rating information has a positive correlation among those with high trust. 
%Therefore, we estimate the total trust matrix and estimate the rating information through the given sparse trust information.
	
	\item \textbf{CDAE}: Collaborative Denoising Autoencoder \cite{Wu2016} used a denoising autoencoder with user latent variables. 
%CDAE is a model with a rating matrix as an input, and it can learn a robust representation by adding noise to the input. 
%This is a generalized version of a latent factor model on CF in deep learning.
	
	\item \textbf{CDL}: Collaborative Deep Learning \cite{Wang2015} used the deep learning and the CF, jointly. CDL improves performance by using document information additionally, and CDL uses SDAE to learn document manifold. 
%CDL couples the deep representations of the document and latent vector of CF through two-way interactions.
\end{itemize}

Appendix E provides detailed specifications for replications of this work, and Appendix F illustrates the sensitivity analysis of $\lambda_{y}$ and $\lambda_{\tau}$.

\subsection{Quantitative Evaluations}
We performed the five-fold cross-validation to quantitatively evaluate the variations of NIPENs, and the performance measures are RMSE, MAE, accuracy, and negative average log-likelihood (NALL) measures. We compared nine models: five baseline models in section 4.2, and four NIPEN variations, which are NIPEN-PGM(SDAE), NIPEN-PGM(VAE,approx.), NIPEN-PGM(VAE), and NIPEN-Tensor. NIPEN-PGM has three variants by choosing either SDAE or VAE as the autoencoder for the text modeling, and by choosing either using the whole matrix for the influence or the low-rank approximated matrix of the influence. 

Table \ref{table:quant1} statistically confirms that the best performance model in every metric is always a variation of NIPEN, which is confirmed with statistical significance. In detail, first, we compare NIPEN-PGM(VAE) and NIPEN-PGM(SDAE), and their performance gap is larger in \textit{Politic2013} than in \textit{Politic2016} which is a relatively sparse setting as shown in Table \ref{table:dataset}. We conjecture that NIPEN-PGM(VAE) is better in handling the sparse dataset than NIPEN-PGM(SDAE). Second, NIPEN-Tensor is a model that considers the correlation between topics, and NIPEN-Tensor may have a better performance when a bill text has multiple topics with complex and rich textual information. As discussed in Section \textit{Datasets on Political Ideal Points}, \textit{Politic2016} has richer textual information than \textit{Politic2013}, and we conjecture that this is the reason why NIPEN-PGM(VAE) in \textit{Politic2013} and NIPEN-Tensor in \textit{Politic2016} show better performances. Third, while the accuracy improvement is relatively small, the improvements on other metrics, particularly RMSE and MAE, are relatively large. Already, the baseline models achieve the accuracy higher than 95\%, so the accuracy improvement could seem minimal. However, our likelihood estimation of \textit{YEA} and \textit{NAY} is considerably improved given the RMSE and the MAE improvement. 

\subsection{Qualitative Evaluations}
\begin{table} 
	\caption{Selected top-five words for each topic. The number of listed topics was set to ten.}
	%\centering
	\begin{center}
		\begin{tabular}{|C{0.2cm}|C{2cm}|C{5cm}|} 
			\hline
			& Topic Label                & Topic Words                                                \\ \hline
			1  & Business and Finance       & Forprofit, Nonrefundable, Govern, SBDC, Financings            \\ \hline
			2  & Disasters Management       & Stabilization, Homeless, Disasters, Alerts, USPS          \\ \hline
			3  & International Relationship & Kuwait, Distributes, Lawsuits, Threatens, Spain            \\ \hline
			4  & Racism                     & Contrary, Black, Compared, Tuskegee, Reagan                 \\ \hline
			5  & Defense                    & United, Soviet, Antiterrorist, IDA, NGA			\\ \hline
			6  & Agriculture                    & Pima, Climate, Cropland, Bush, Badlands			\\ \hline
			7  & Social                    & Contribute, Donors, Childcare, Resettlement, DRR	\\ \hline
			8  & Health                    & FEHBP, Heroin, Stability,
			Musculoskeletal, Transplantation			\\ \hline
			9  & Foregin                    & Agency, Lantos, FPI,
			fusion, division			\\ \hline
			10  & International Trade  & Clearinghouses, ESF,  Discrepancies, Repay, Charging	\\ \hline
		\end{tabular}
	\end{center}
	\label{table:word_topic}
\end{table}
\begin{figure}[h!]
	\centering
	\includegraphics[width=2.8in]{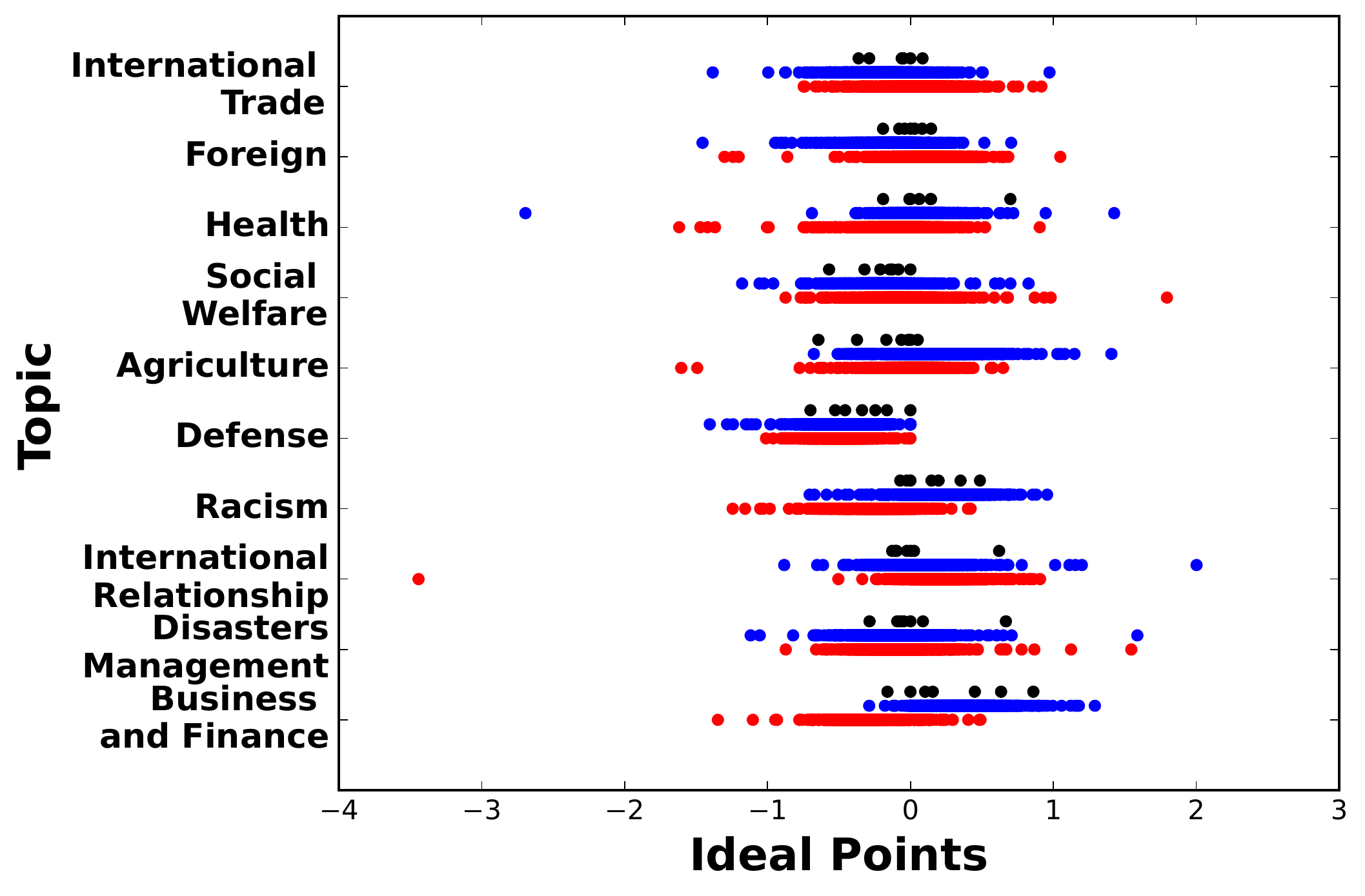}\hspace{0em}%	
	%	\smallskip
	\caption{\label{fig:x_uk}Individual legislators' ideal points for each topic}
\end{figure}

\begin{figure}[h!]
	\centering
	\includegraphics[width=2.6in]{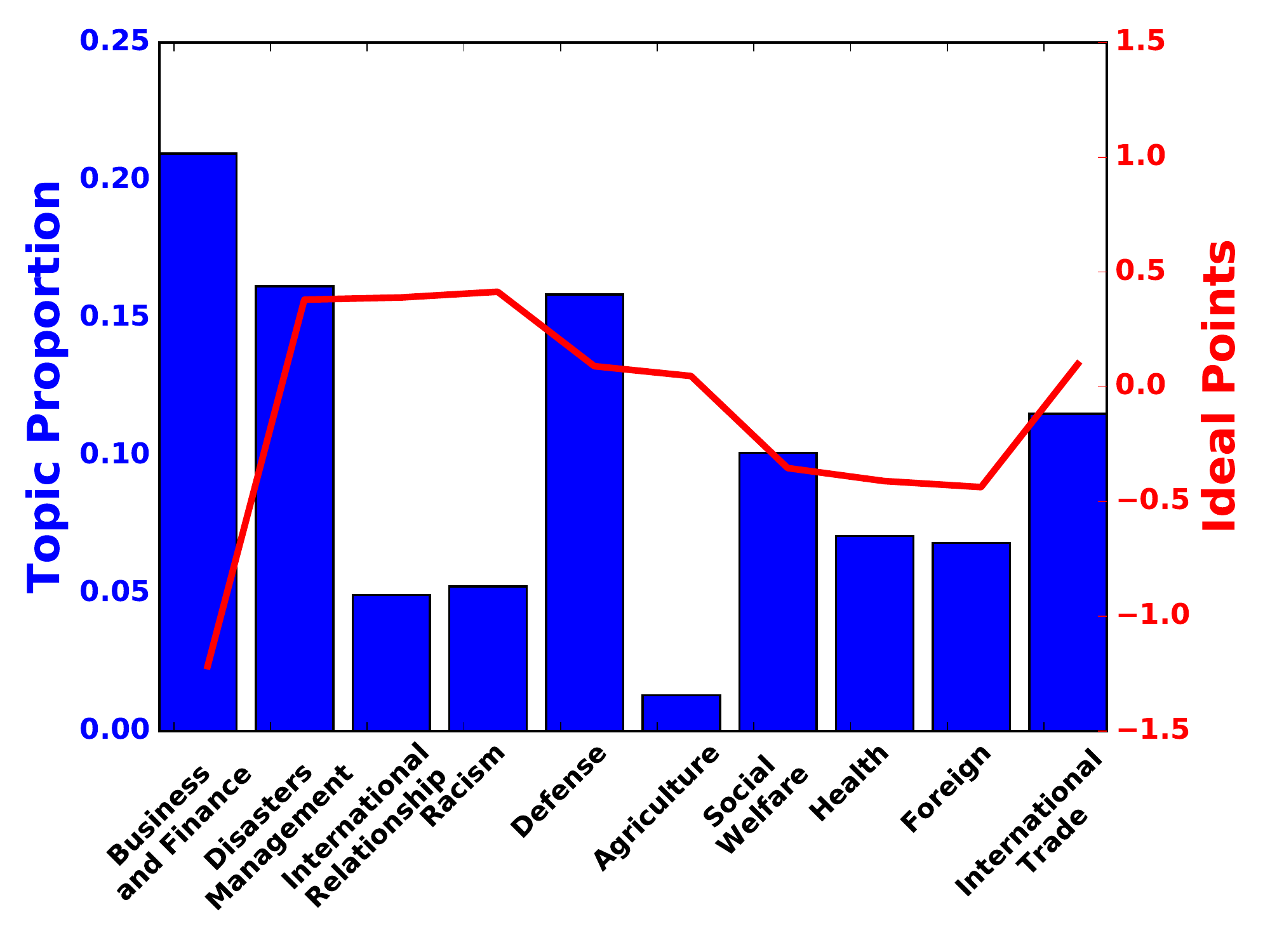}\hspace{0em}%	
	%	\smallskip
	\caption{\label{fig:z_dk_a_dk}Topic proportion and ideal points of H.Res.794 (114th) bill}
\end{figure}

In addition to the quantitative results, we interpret the latent variables of NIPEN-PGM(VAE) on \textit{Politic2016}. First, to comprehend the dataset and the qualitative results, we computed the word-topic matrix from well-learned VAE variables, $\psi_{1}$, as shown in Table \ref{table:word_topic}. This table provides a snapshot of topics in the bills. Then, we relate this topic to the bill ideal points, $a_{dk}$. The latent dimension, $k$, becomes the common dimension of an ideal point value and a topic weight for each topic in the bill. Figure \ref{fig:z_dk_a_dk} shows an example of the topic weight as the bar chart and the ideal point value as the line chart. The illustrated bill, or H.Res.794 (114th), has the largest absolute value, $|a_{dk}\widetilde{z}_{dk}|$ in a 'Business and Finance' topic where $\widetilde{z}_{dk}$ denotes the normalized $z_{dk}$. 

This bill ideal point is correlated with the legislator ideal point, $x_{uk}$, to generate the vote records. Here, the dimension, $k$, is the same latent dimension of the topic in Table \ref{table:word_topic}, and we provide the scatter plot of the legislators' ideal points per topic in the Figure \ref{fig:x_uk}. The prior mentioned bill (H.Res.794 (114th)) considers the appropriations for financial services and general government, and the major topic is \textit{Business and Finance}, and the bill ideal point in \textit{Business and Finance} is -1.217. Together, the vote casting will be determined by the legislators' view on \textit{Business and Finance}, and this topic shows the greatest disagreement between the Republicans and the Democrats according to the Figure \ref{fig:x_uk}. In the real world, the voting results were same as expected: 1) the voting was very partisan, 92.2\% Republican voted \textit{YEA} and the 90.3\% Democrat voted \textit{NAY}.
\begin{figure}[t!]
	\centering
	\includegraphics[width=2.8in]{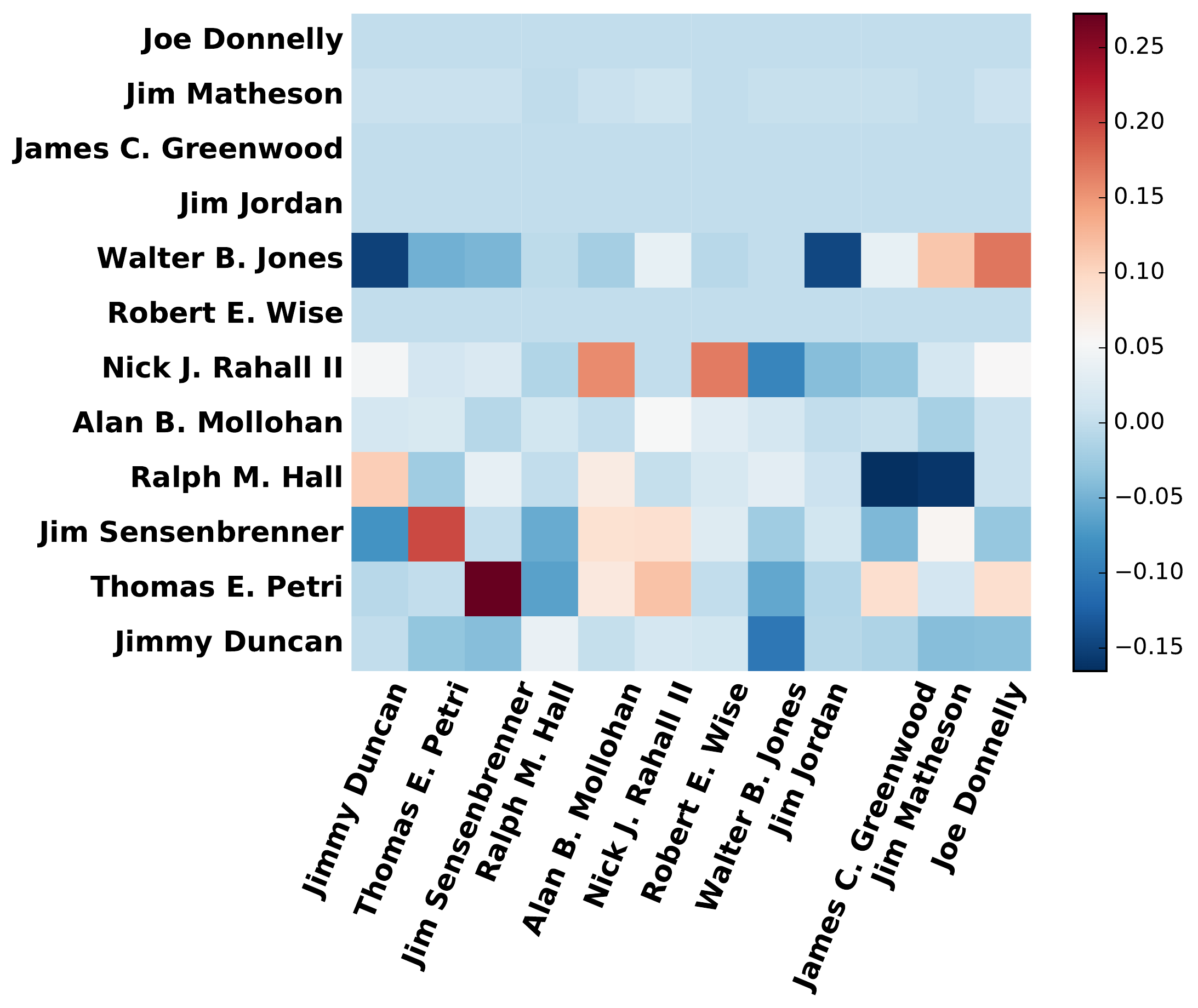}
	\caption{Trust network between legislators}
	\label{fig:tau_uu}
\end{figure}

The second qualitative interpretation focuses on the legislators' network. We selected 12 legislators who have either strongly positive or negative relationships with each other, shown in the Figure \ref{fig:tau_uu}. In general, the legislators have a strong positive relationship when they have the same district and the party. Among the top-five positive relationships, four of them have the same party and the same district, i.e. 'Thomas E. Petri$\leftrightarrow$Jim Sensenbrenner', 'Nick J. Rahall II$\rightarrow$Robert E. Wise', and  'Nick J. Rahall II$\rightarrow$Alan B. Mollohan'\footnote{$\tau_{uu}$ is asymmetric matrix. arrow('$\rightarrow$') indicates the direction of the trust}. The closest relations are 'Thomas E. Petri' and 'Jim Sensenbrenner'. They were both republican representatives from Wisconsin, and they share similar voting patterns. They have voted 6,288 times for the same bill, and the 5,764 votes were same ($91.6\%$). Especially, they voted \textit{NAY} for H.R.730 (111th) which is a "suspension of the rules", and 397 legislators votes \textit{YEA}. For H.R.6063 (110th), 'Thomas E. Petri' and 'Jim Sensenbrenner' voted \textit{NAY} together while 94.4$\%$ legislators voted \textit{YEA}. We report further analyses in Appendix G.

\begin{figure}[h!]
	\centering
	\includegraphics[width=2.5in]{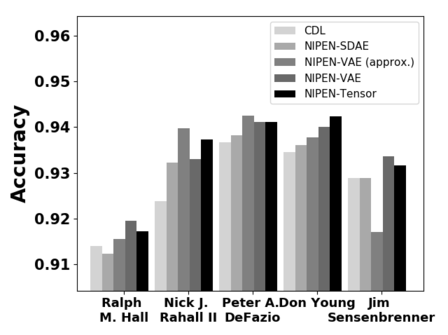}
	\caption{Accuracy of top five legislators who are affected by network factor}
	\label{fig:network_accuracy}
\end{figure}

The third qualitative analysis concentrates on the interaction between the contents and the network parts. We used two scaling variables $\alpha_{u}$ and $\beta_{u}$, which controls the strengths of contents factor and network factor, respectively. Table \ref{table:contents_network} shows the top-five legislators who were affected by either contents or network factors. Since the variations of NIPEN is an integrated model of network modeling as well as the textual bill modeling, the NIPENs should better perform than the baseline models, i.e. CDL, which only models the texts, and Figure \ref{fig:network_accuracy} confirms this hypothesis.

\begin{table}
	\caption{Top-five legislators who are affected by contents or network factors a lot. The scaling variable ($\alpha_{u}$ for contents based, and $\beta_{u}$ for network based), political party, and district of the member are indicated in parentheses.}
	%\centering
	\begin{center}
		\begin{tabular}{|C{0.3cm}|C{3.2cm}|C{3.2cm}|} 
			\hline
			{\textbf{}} & 
			{\textbf{Contents based}} & {\textbf{Network based}} \\ \hline
			
			\multirow{2}{*}{1} & Ron Paul & Ralph M. Hall \\
			& (0.260, R, TX) & (0.304, R, TX) \\ \hline
			
			\multirow{2}{*}{2} & Virgil H. Goode & Nick J. Rahall II \\
			& (0.220, R, VA) & (0.250, D, WV) \\ \hline
			
			\multirow{2}{*}{3} & Dennis J. Kucinich & Peter A. DeFazio \\
			& (0.218, D, OH) & (0.247, D, OR) \\ \hline
			
			\multirow{2}{*}{4} & Henry Cuellar & Don Young \\
			& (0.198, D, TX) & (0.228, R, AK) \\ \hline
			
			\multirow{2}{*}{5} & Walter B. Jones & Jim Sensenbrenner. \\
			& (0.195, R, NC) & (0.227, R, WI) \\ \hline
		\end{tabular}
	\end{center}
	\label{table:contents_network}
\end{table}

\section{Conclusion}
We proposed two versions of machine learning models, NIPEN-PGM and NIPEN-Tensor, to analyze the ideaology in the legislation process. 
%NIPEN-PGM is a hybrid model of a probabilsitc graphical model of the voting process and the network influence; and a neural network model of the document latent modeling. NIPEN-Tensor is a generalized version of NIPEN-PGM by modeling the cross interactions between inferred topics. 
The variations of NIPEN show the state-of-the-art performance in all measures on \textit{Politic2013} and \textit{Politic2016}. Furthermore, NIPEN provides various interpretations in why \textit{YEA} or \textit{NAY} is casted by illustrating 1) the ideal point estimation of individual legislators and bills; 2) the trust network between legislators; and 3) the content and network influence for each legislator. These supervised and unsupervised tasks could be critical insights into quantitatively understanding politics in the legislative process.

\title{Appendix: Neural Ideal Point Estimation Network}
%\author{}
\maketitle

\section{Appendix A. Problem Formulation}
In general, the influence on legislative voting originates from 1) the individual ideal points of the legislator, 2) the contents of the bill, and 3) the interests of a political group that a legislator belongs to. We operationalize these influence structure as the concepts defined in the below.

\begin{defn}
	\textbf{Ideal point} is a measure of legislator's preference for each topic when we have $K$ topics in our bill texts. The ideal point for a particular topic $k$ of a particular member $u$ is represented by $x_{uk}$, and it follows $N(0,\lambda_{u}^{-1})$. The sign of $x_{uk}$ represents the preferred voting direction (positive or negative), and the size of $|x_{uk}|$ represents the preference strength. The ideal point for a particular topic $k$ of a particular bill $d$ is represented by $a_{dk}$, and its distribution is $N(0,\lambda_{u}^{-1})$. The interprestation of $a_{dk}$ is same as $x_{uk}$.
\end{defn}

\begin{defn}
	\textbf{Contents} refer to the bill elements, i.e. text descriptions, which affect the voting result. The latent representation of the contents is $y_{dk}$ which is the addition of $z_{dk}$, the topic of the bill; and $\xi_{dk}$, the deviation of the bill from the topic of the bill text.
\end{defn}

\begin{defn}
	\textbf{Network} means the collection of relationships between legislators whose vote affect the other's vote.
	The strength of network relationships is modeled as $\tau_{uu'}$, which follows $N(0,\lambda_{\tau}^{-1})$.
	The sign of $\tau_{uu'}$ indicates the voting alignment between $u$ and $u'$ legislators, and $|\tau_{uu'}|$ means its alignment strength.
	This study assumes that the network relationships are asymmetric bidirectional, and only the legislators in the same term affect each other.
\end{defn}

\begin{defn}
	\textbf{Scaling parameters} mean the influence of contents and networks when a legislator votes. $\alpha_{u}$ is a content scaling parameter, and $\beta_{u}$ is a network scaling parameter. Each scaling parameter is a $|U|$-dimensional vector, followed by $N(0,\lambda_{\alpha}^{-1})$. This study assumed that the degree of influence on the contents and the network would be different per each legislator.
\end{defn}

Now, given the above defined concepts, we enumerates the research questions to test with NIPEN. 
\begin{prob}
	NIPEN can predict the results of the voting by inferring the bill topic, the bill ideal points, the legislator ideal points, and the network relationships between the legislators.	
\end{prob}

\begin{prob}
	NIPEN provides the interpretation on the voting results of the bill. For example, NIPEN illustrates the interpretable latent information from the bill topic, the bill ideal point, and the legislator ideal point taking into account the correlation between the topics.
\end{prob}

\begin{prob}
	NIPEN can 1) analyze the trust between legislators (individual unit), and 2) the trust network comparison between parties (group unit).
\end{prob}

\begin{prob}
	NIPEN provides the behavioral analyses on legislators from the voting motivation perspective, which could be motivated by either legislator ideal point or network relationship.
\end{prob}

\section{Appendix B. Document Modeling Autoencoders}

\subsection{Appendix B.1. Variational Autoencoder (VAE)}
NIPEN extracts the topics of the legislative bills with VAE \cite{Kingma2013} which is a type of deep generative model. VAE learns the disentangled and low-dimensional representation of high dimensional data through the probabilistic encoding, or $q_{\phi}(z|w)$; and the probabilistic decoding, or $p_{\theta}(w|z)$. Therefore, the original objective function of VAE is composed of the linear sum of corresponding two terms. The first term originating from the encoding is the KL divergence between the probabilistic encoding and the prior for latent variable, or $p_{\theta}(\mbz)$; and this term enforces the regularization. The second term is the expectation on the negative reconstruction error, and this term is related to the decoding part.  By putting both terms together, the objective function follows as Eq. (\ref{eq:VAE_loss_1}).
\begin{equation}  \begin{aligned} \label{eq:VAE_loss_1}
		\cL(\theta,\phi) = -{\operatorname{D_{KL}}}(q_{\phi} (\mbz | \mbw)\|p_{\theta}(\mbz)) + \mathbb{E}_{q_\phi}[\log p_{\theta}(\mbw | \mbz)] 
\end{aligned} \end{equation}

Given the high variance over $\phi$, a direct optimization of Eq. (\ref{eq:VAE_loss_1}) is not efficient. Hence, Kingma and Welling (2014) suggested the re-parametrization trick as follows: 1) Draw $\epsilon^{l} \sim N(0,I)$, and 2) Optimize the mean and standard deviation of $q_{\phi}(z|w)$, and 3) Compute $z^{l} = \mu(w) + \sigma(w) \odot \epsilon^{l}$. From the trick, the objective function of VAE is turned into Eq. (\ref{eq:VAE_loss_2}) where $L$ is the number of samples.
\begin{equation}  \begin{aligned} \label{eq:VAE_loss_2}
		\widetilde{\cL}(\theta,\phi) = -{\operatorname{D_{KL}}}(q_{\phi} (\mbz | \mbw)\|p_{\theta}(\mbz)) + \frac{1}{L}\sum_{l=1}^{L} \log p_{\theta}(\mbw | \mbz^{l}) 
\end{aligned} \end{equation}

\subsection{Appendix B.2. Stacked Denoising Autoencoder (SDAE)}
NIPEN-PGM(SDAE) extracts the topics of the legislative bills with SDAE \cite{VincentPASCALVINCENT2010}. SDAE learns the disentangled latent feature through the encoding and decoding with bottleneck and corrupted input. SDAE use the corrupted input $\mbw_{c}$ instead of the original input $\mbw$ to force the relationship learning. The SDAE is optimized for the purpose of reconstructing the $\mbw$ resulting from the encoding process($f_{e}$) through the encoder weight ($W^{(e)}$) and the decoding$f_{d}$ result through the decoder weight ($W^{(d)}$). The objective function follws as Eq. (\ref{eq:SDAE_loss_1}) 
\begin{equation}  \begin{aligned} \label{eq:SDAE_loss_1}
		\left \| f_{d}(f_{e}(\mbw,W^{(e)}),W^{(d)})-\mbw_{c}) \right \|_{2}^{2} \\
\end{aligned} \end{equation}

\section{Appendix C. Notations}
Table \ref{table:notation} summarizes all symbols used in this study.

%\begin{center}
%\begin{tabular}{|C{0.3cm}|C{3.2cm}|C{3.2cm}|} 
\begin{table}[h!]
	%\centering
	\caption{Notation description}
	\label{table:notation}
	\begin{center}
		\begin{tabular}{|C{1.7cm}|L{6cm}|}
			\hline
			Symbol & Description \\ \hline \hline
			$D$ &         Set of bills  \\ \hline
			$V$ &         Set of Unique words   \\ \hline
			$U(=U')$ &	  Set of legislators	\\ \hline
			$K$ &         Set of topics   \\ \hline
			$G$ &  Rank of $\widetilde{\tau_{1}}$ and $\widetilde{\tau_{2}}$ \\ \hline
			$I_{u}$ & 	Other legislators within the same term as $u$ \\ \hline
			$w_{dv}$ &  Frequency of $v_{th}$ token in document $d$ \\ \hline
			$z_{dk}$ &  Topic proportion for each bill and topic \\ \hline
			$y_{dk}$ &  Bill latent vector  \\ \hline
			$a_{dk}$ &  Ideal point for each bill and topic \\ \hline
			$\eta_{d}$  & Constant offset for each bill $d$      \\ \hline
			$r_{ud}$ & Voting record from legislator $u$ to bill $d$    \\ \hline
			$x_{uk}$  & Ideal point for each legislator and topic \\ \hline
			$\alpha_{u}$  & Contents scaling parameter for legislator $u$ \\ \hline
			$\beta_{u}$  & Network scaling parameter for legislator $u$   \\ \hline
			$\tau_{uu'}$ & Trust network between legislator $u$ and $u'$\\ \hline
			$\widetilde{\tau_{1}}$,$\widetilde{\tau_{2}}$ & Approximated matrix of $\tau_{uu'}$ \\ \hline
			$\xi$  &     Latent offset between $z_{dk}$ and $y_{dk}$ \\ \hline
			$\epsilon$  &     Random noise vector drawn from $N(0,I)$ \\ \hline
			$\phi (\theta)$  &     Parameter of encoder (decoder) in VAE \\ \hline
			$C (N)$  &   Contents (Network) information \\ \hline
			$E$  &   The tensor combined with $x_{uk},y_{dk},a_{dk}$ \\ \hline
			$W^{(T)}, b^{(T)}$  &   Neural tensor network parameter \\ \hline
		\end{tabular}
	\end{center}
\end{table}
% $\widetilde{\tau_{1}}$ $\in  \mathbb{R}^{U \times G}$, $\widetilde{\tau_{2}}$ $\in

\section{Appendix D. Dataset Descriptions}

The first dataset is \textit{Politic2013}, and it was collected from \textit{THOMAS}\footnote{http://thomas.loc.gov/home/rollcallvotes.html}, and \cite{Gu2014}. For an additional experiment, and for more up-to-date analyses, we collected a new roll-call dataset, \textit{Politic2016} from GovTrack .GovTrack provides raw roll-call data, so we processed the expanded part of \textit{Politic2016}, manually. For the research community, we released the code and dataset on https://github.com/gtshs2/NIPEN

\section{Appendix E. Experiment Settings}

For TFIPM, we followed the optimal parameters that the author reported. We set the latent dimension($K$), the trade-off weight, and the regularization weight as 10, 0.8, and 22.4, respectively. The latent dimension of CDL and NIPENs was set to ten, equally. For Autorec, the optimal number of the latent dimension and the regularization parameter are 100 and 0.001, respectively. Trust SVD shows the best performance when the weights of CF and trust part are 1,000 and 0.001, respectively, while $K$ is set to 10. For CDAE, we set the regularization weight, the corruption ratio, and the number of latent dimension as 0.001, 0.4, and 50, respectively; and the encoder and the decoder activation functions are sigmoid. For CDL, we find that the optimal parameters of $\lambda_{u}$,$\lambda_{v}$,$\lambda_{w}$,$\lambda_{n}$, the dropout rate and the activation functions are 0.01, 100, 1, 100, 0.1 and the sigmoid function, respectively. The neural network structure of Autorec, CDAE, CDL, NIPENs are set to [512,128,$K$,128,512], equally. Finally, we set the parameters of NIPENs such as $\lambda_{f}=10$, $\lambda_{y}=10$, $\lambda_{u}=0.1$, $\lambda_{\tau}=1$, $\lambda_{\alpha}=1$, $\lambda_{n}=1000$ $G=3$, and we performed grid searches to find the optimized parameters of NIPENs. Finally, NIPEN-Tensor has the two-layered tensor $E$ in the topic axis.

\section{Appendix F. Hyperparameter Study}

\begin{figure}[h!]
	\centering
	\includegraphics[width=1.6in,height=1.25in]{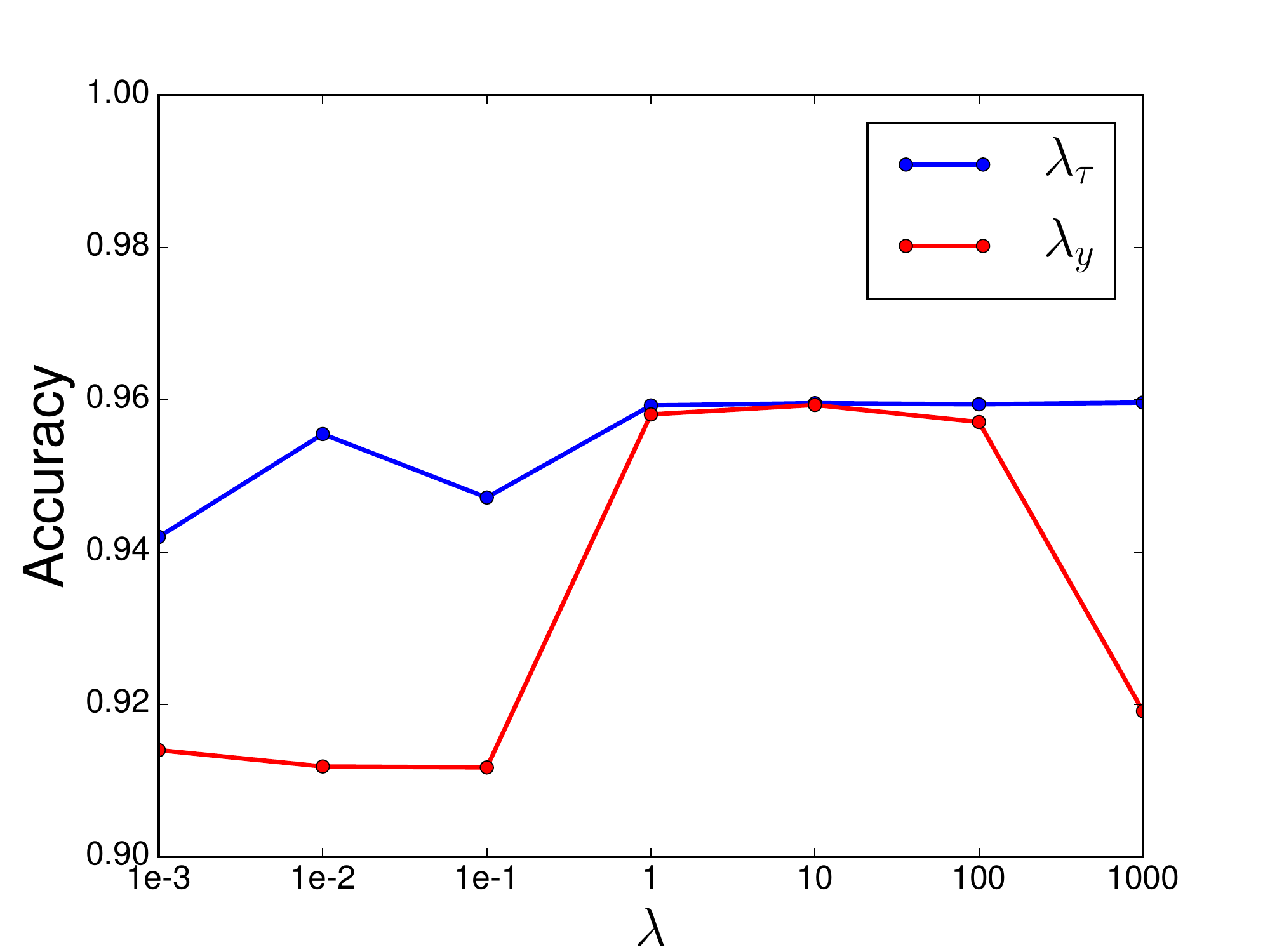}\hspace{0em}%
	\includegraphics[width=1.6in,height=1.25in]{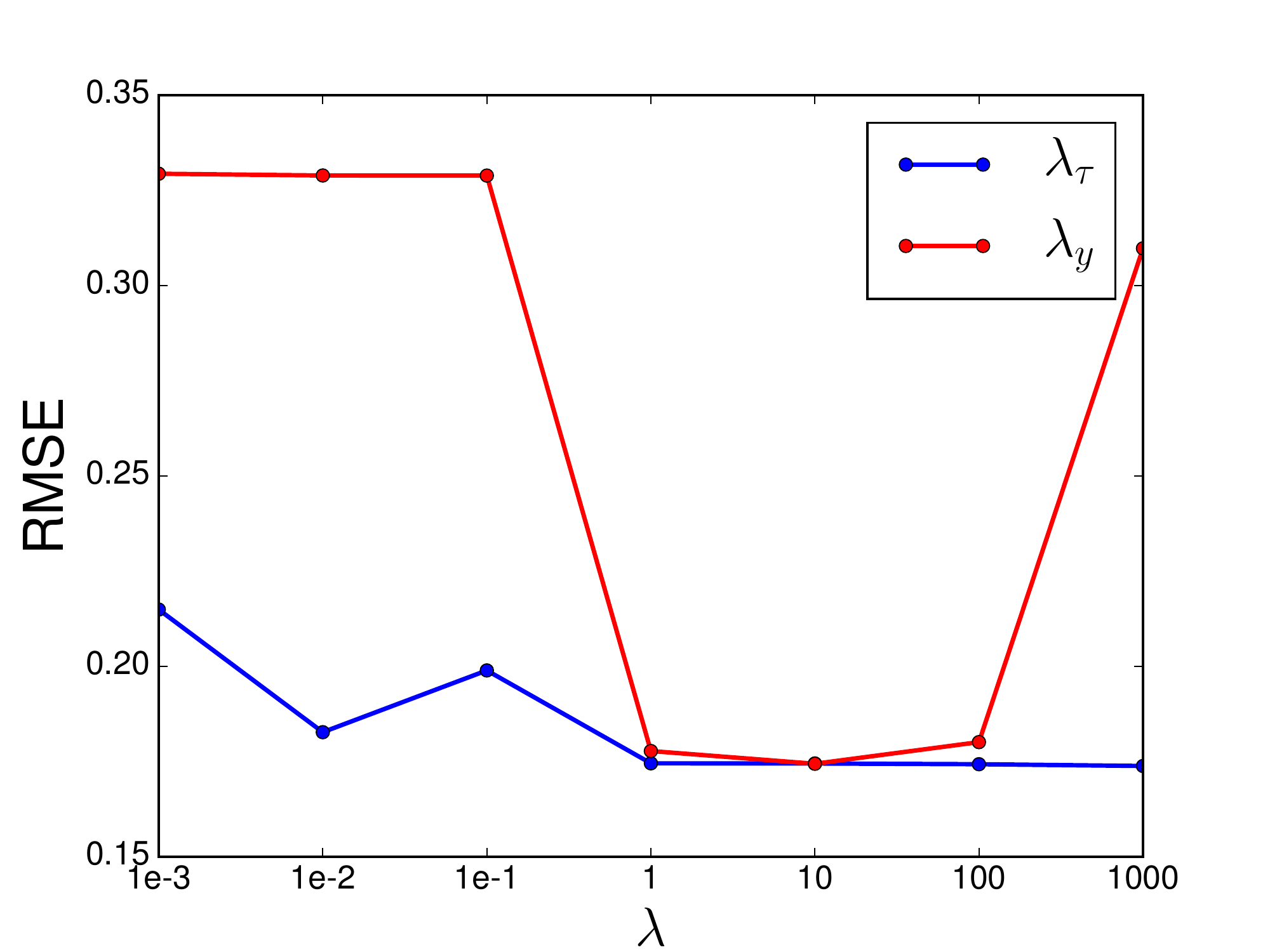}
	\smallskip
	\includegraphics[width=1.6in,height=1.25in]{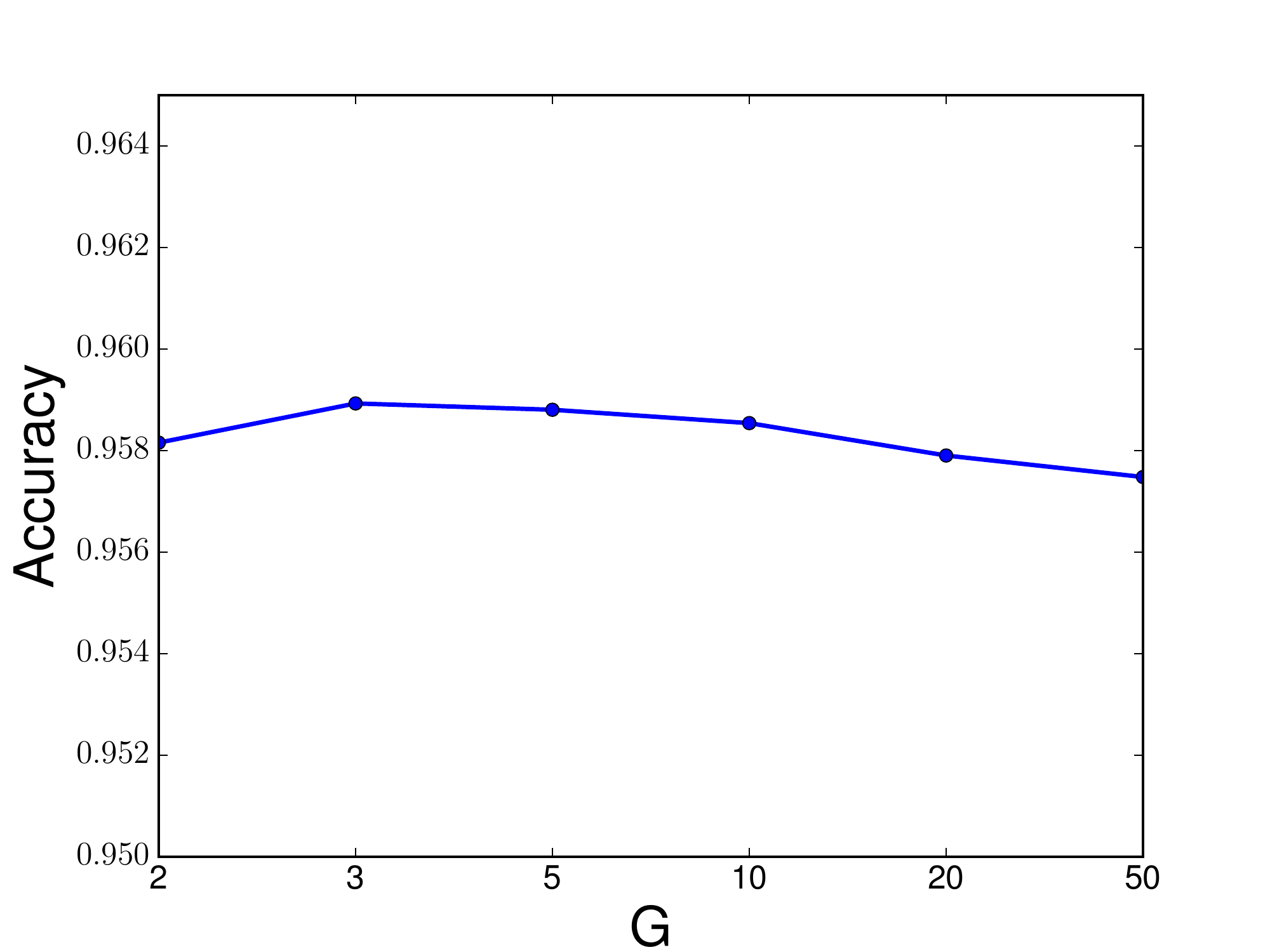}\hspace{0em}%
	\includegraphics[width=1.6in,height=1.25in]{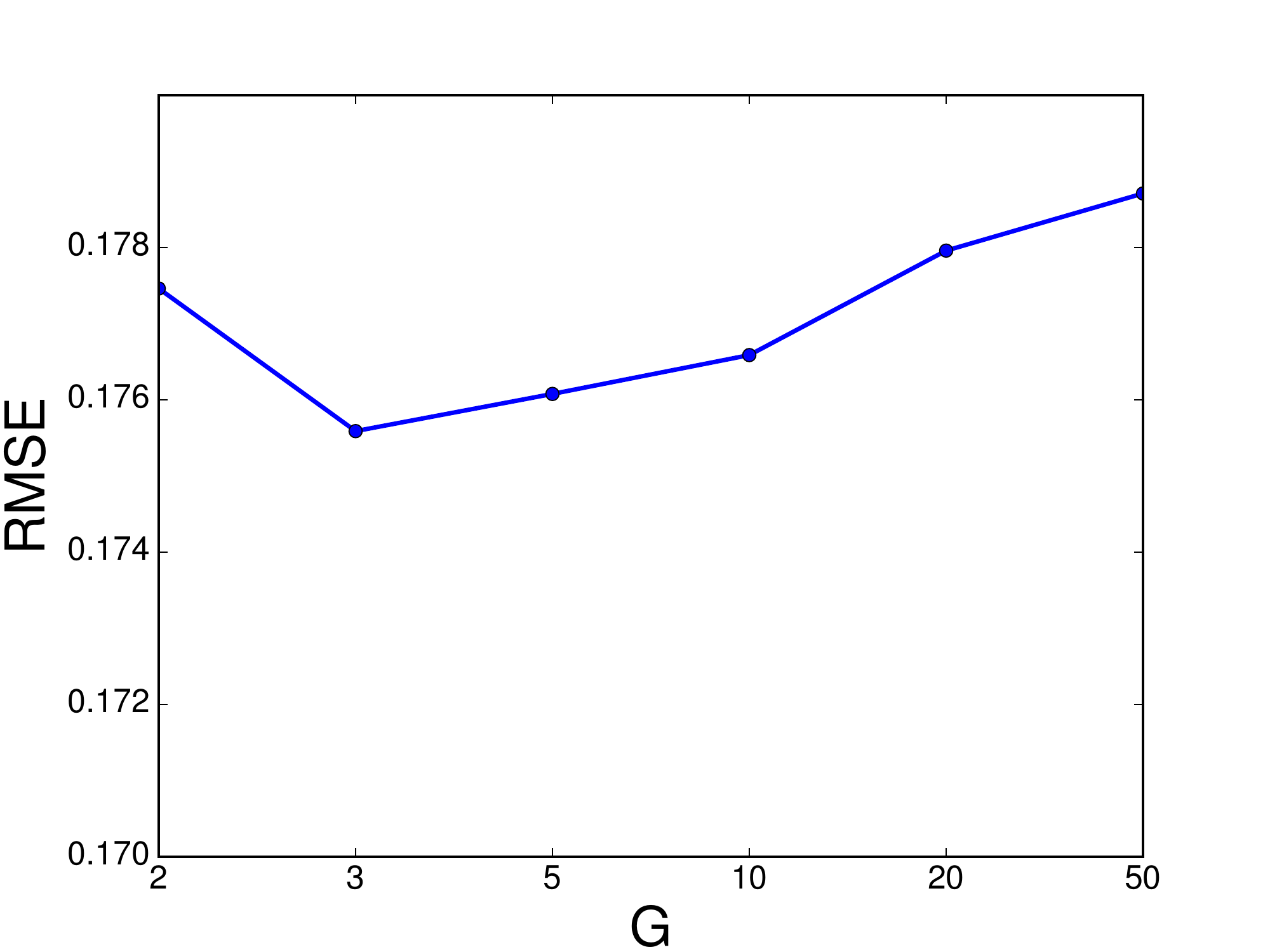}
	
	\caption{\label{fig:parameter_study}(Top-Left) Accuracy for $\lambda_{y}$ and $\lambda_{\tau}$ value, (Top-Right) RMSE for each $\lambda_{y}$ and $\lambda_{\tau}$ value, (Bottom-Left) Accuracy for $G$ value, (Bottom-Right) RMSE for each $G$ value}
\end{figure}

This quantitative improvement requires a well-tuned hyperparameter setting, illustrated in Figure \ref{fig:parameter_study}. $\cL_{NIPEN}$ enumerates multiple hyperparameters, and we found that $\lambda_{y}$ and $\lambda_{\tau}$ are the most important parameters to decide. $\lambda_{y}$ specifies the causality strength from the bill text latent in VAE to the bill latent in the legislative CF. The low value of $\lambda_{y}$ will separate VAE and CF, but its high value will disrupt the manifold learning of VAE. Moreover, $\lambda_{\tau}$ specifies the regularization strength from the legislators' network influence to the voting. The small value of $\lambda_{\tau}$ will overfit the network influence, and its large value will limit the learning of network influence model in CF. 

\section{Appendix G. Results in Network Influence Analyses}

\begin{figure}[h!]
	\centering
	\subcaptionbox{Threshold = 0.03\label{fig:repdem003}} {\includegraphics[width=1.6in]{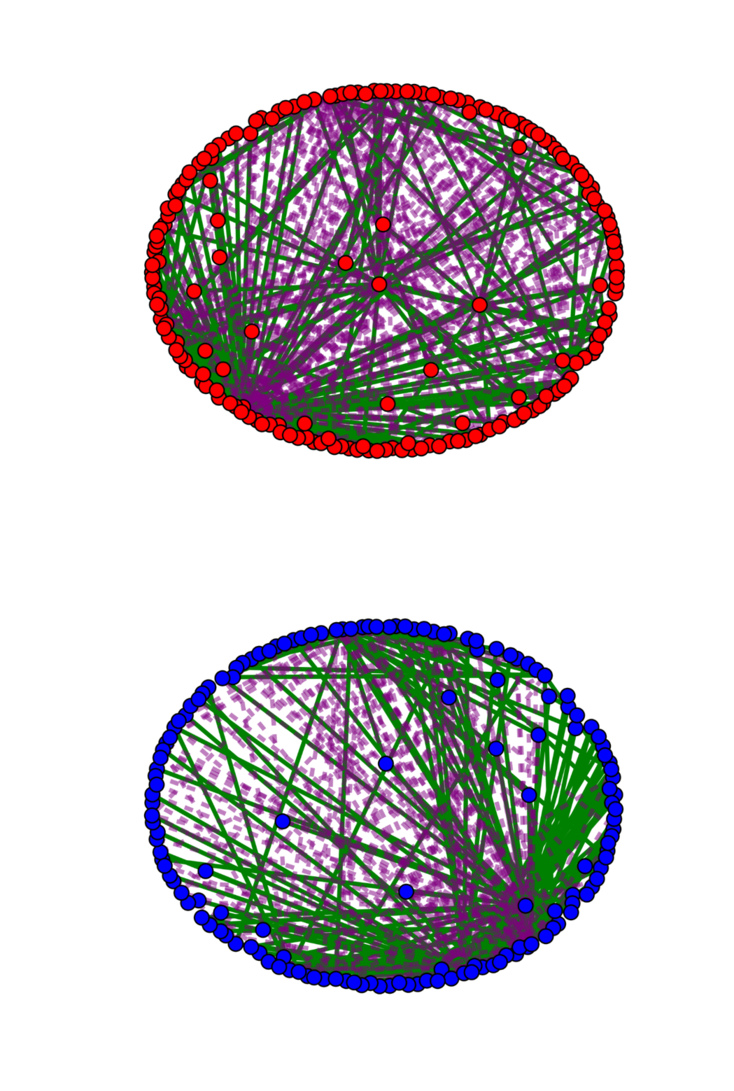}}\hspace{0em}%
	\subcaptionbox{Threshold = 0.05\label{fig:repdem005}} {\includegraphics[width=1.6in]{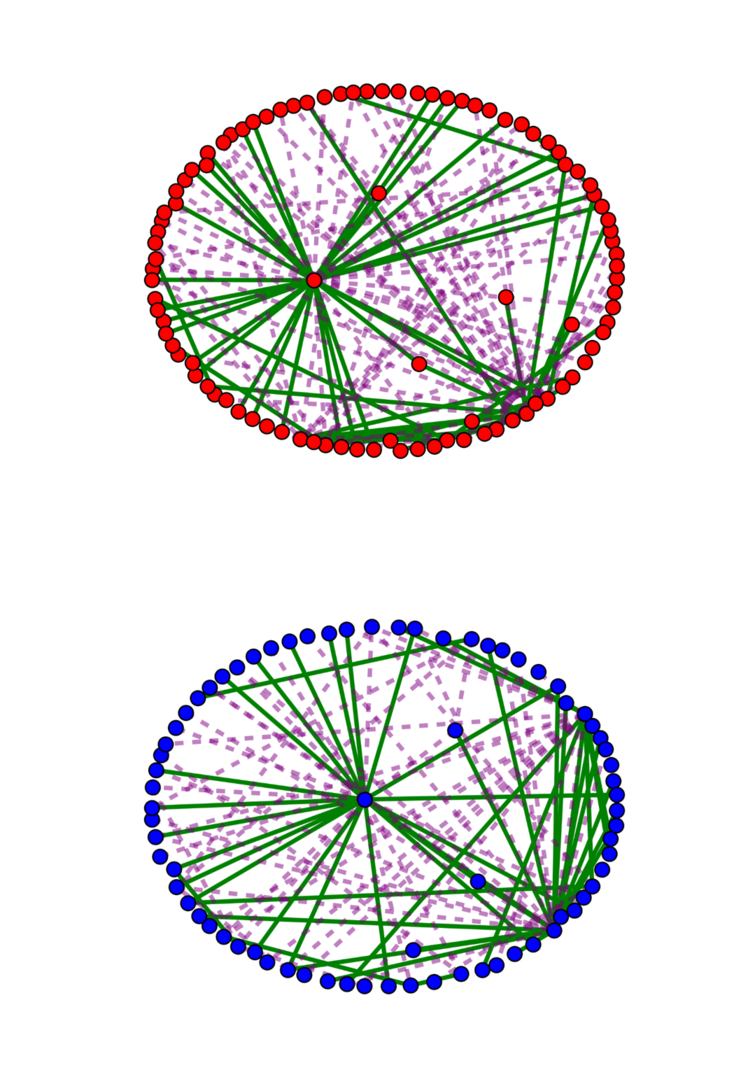}}
	\subcaptionbox{Threshold = 0.07\label{fig:repdem007}} {\includegraphics[width=1.6in]{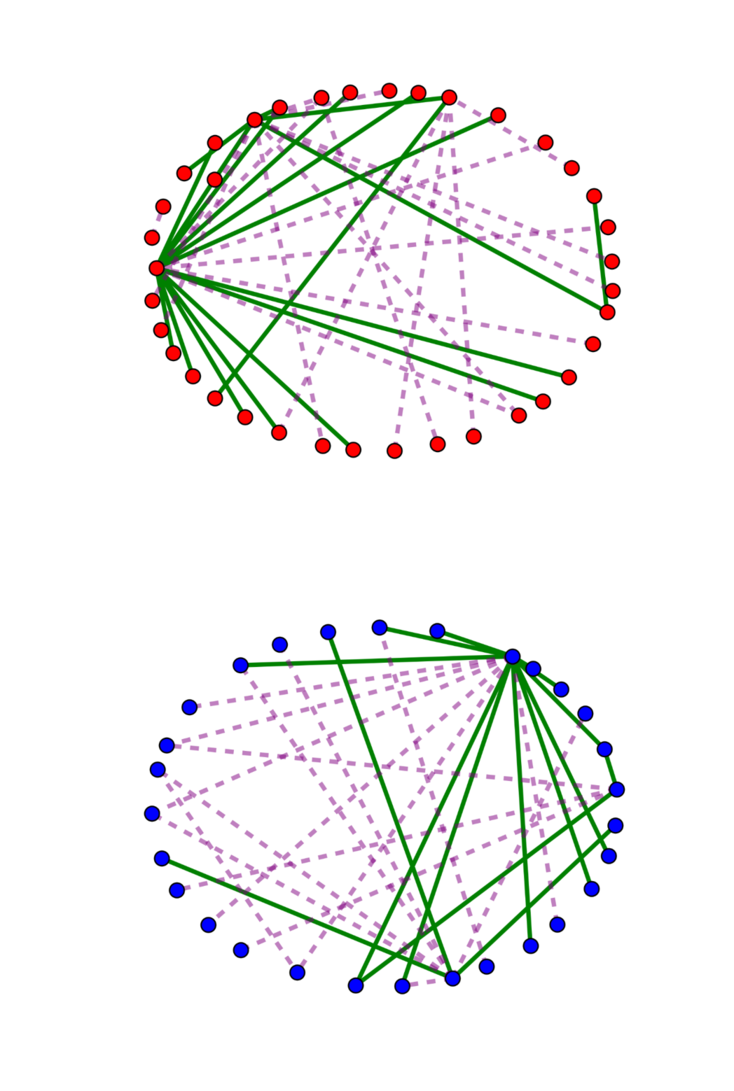}}\hspace{0em}%
	\subcaptionbox{Threshold = 0.1\label{fig:repdem010}} {\includegraphics[width=1.6in]{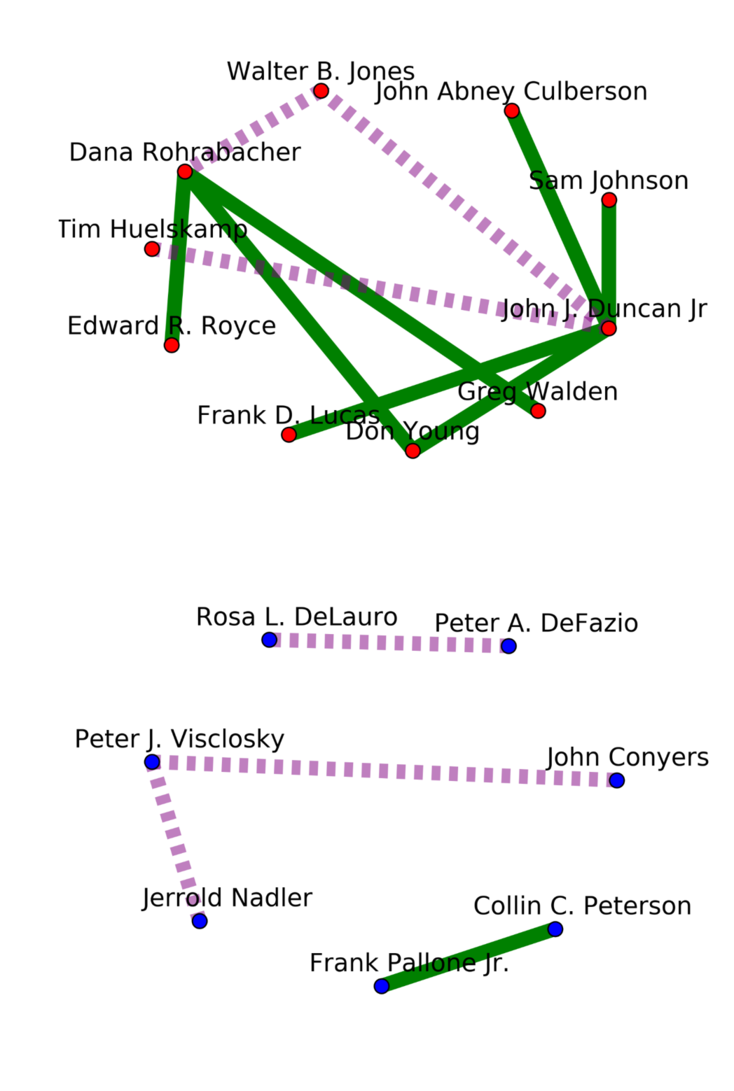}}
	
	\caption{Network influence visualization within each party in December 2016. The top visualizations represent the Republican Party, and the bottom visualizations represent the Democratic Party. If the network influence, $\tau_{uu'}$, between two legislators exceeds the threshold, the relationship is expressed as a green solid line. If the network influence, $\tau_{uu'}$, is less than the threshold, the relationship is noted as a purple dotted line.}
	\label{fig:total_network_threshold}
\end{figure}

To examine the network relationships within each party, Figure \ref{fig:total_network_threshold} illustrates the trust network parameter. A red node represents a Republican; a blue node stands for a Democrat; a green solid line indicates an inferred close relationship; and a purple dotted line indicates an inferred unfriendly relationship. Taking a threshold at 0.1 and looking at $\tau_{uu'}$ with values greater than 0.1 and less than -0.1 (Figure \ref{fig:repdem010}), John J. Duncan Jr and Dana Rohrabacher have the greatest network impact given their number of connected legislators in the Republican party. The commonalities between the two influential legislators are 1) being a member of the House of Representatives; and 2) having been politically active for a long time (Duncan started as a congressman in Tennessee in 1988 and Laura Baker as a California congressman in 1989. Especially, Jimmy Duncan is the House's longest-serving Republicans.).

We compared and contrasted the network structure of Republicans and Democrats. As shown in Table \ref{table:party_network_comparison}, the network influence among the total members is greater in the Democratic Party given its mean value of $|\tau_{uu'}|$. However, the Republican party has the higher number of influential legislators when we limit the network influence with thresholds, i.e. when we limit $|\tau_{uu'}| > 0.05$ or $|\tau_{uu'}| > 0.1$ in Table \ref{table:party_network_comparison}. This suggests that the Democrats have averagely higher and more equal influences between the members while the Republicans have a number of authorative influencers among the members.

\begin{table}[]
	\centering
	\caption{Comparison of network influence by each Party}
	\label{table:party_network_comparison}
	\begin{tabular}{|C{3.8cm}|C{1.7cm}|C{1.7cm}|}
		\hline
		& Republican & Democratic \\ \hline
		\# of pair s.t. $\tau_{uu'} > 0.05$ &   58     &    49     \\ 
		\hline
		\# of pair s.t. $\tau_{uu'} < -0.05$ &     67     &   53   \\ 
		\hline
		\# of pair s.t. $\tau_{uu'} > 0.1$ &     7     &  1    \\ 
		\hline
		\# of pair s.t. $\tau_{uu'} < -0.1$ &    3    &     3   \\ 
		\hline
		Mean of $|\tau_{uu'}|$ &  9.0E-04   &   0.0018   \\ 
		\hline
		Variance of $|\tau_{uu'}|$ &  2.49E-05   &  3.38E-05  \\ 
		\hline
	\end{tabular}
\end{table}

\begin{table}[h!]
	\centering
	\caption{Content and Network scale comparison between the Party. Average Ranking@$K$ is the average of the ranking of each legislator, by the party, when the ranking is made up to the top $K$ legislators, based on scaling parameter ($\alpha_{u}$ $\beta_{u}$).}
	\label{table:scaling_by_party}
	\begin{tabular}{|C{3.2cm}|C{0.8cm}|C{0.8cm}||C{0.8cm}|C{0.8cm}|}
		\hline
		& \multicolumn{2}{c||}{Content ($|\alpha_{u}|$)} & \multicolumn{2}{c|}{Network ($|\beta_{u}|$)} \\ \hline
		& R    & D    & R    & D   \\ \hline
		NL@10  & 6   & 4     & 6   & 4    \\ \hline
		NL@100 & 40   & 60    & 42    & 58  \\ \hline
		AR@100   & 52.87   & 48.91  & 46   & 53.75  \\ \hline
		AR@All & 774.6  & 762.7 & 781.4 & 754.9  \\ \hline
		Mean value of $\alpha_{u}$ ($\beta_{u}$) & 0.149  & 0.148   & 0.068  & 0.064 \\ \hline
	\end{tabular}
	\begin{tablenotes}
		\small
		\item R : Republican Party / 	D : Democratic Party
		\item AR@K : Average Ranking at top K
		\item NL@K : Number of legislators at top K
	\end{tablenotes}
\end{table}

The United States has a two-party system. In order to analyze this reality effectively, contents scaling parameter and network scaling parameter are analyzed by each party.
Comparing the mean value of $|\alpha_{u}|$ and $|\beta_{u}|$ in the Table \ref{table:scaling_by_party}, we can see that both parties are generally voting rather than networks, concentrating on their own politics and the contents of the bill itself.
To compare two political parties relative to each other based on Average Ranking@100 in Table $\ref{table:scaling_by_party}$, the Republican Party generally relies on the network, and the Democratic Party on contents, when they are voting.
If we associate this conclusion with Table $\ref{table:party_network_comparison}$, we can explain that a Democratic have more influence in the network within the party. However, the proportion of people who are dependent on the network (including network from their own party and opposing party) is relatively high in the Republican Party.

\begin{figure}[h!]
	\centering
	\includegraphics[width=3.2in,height=2.5in]{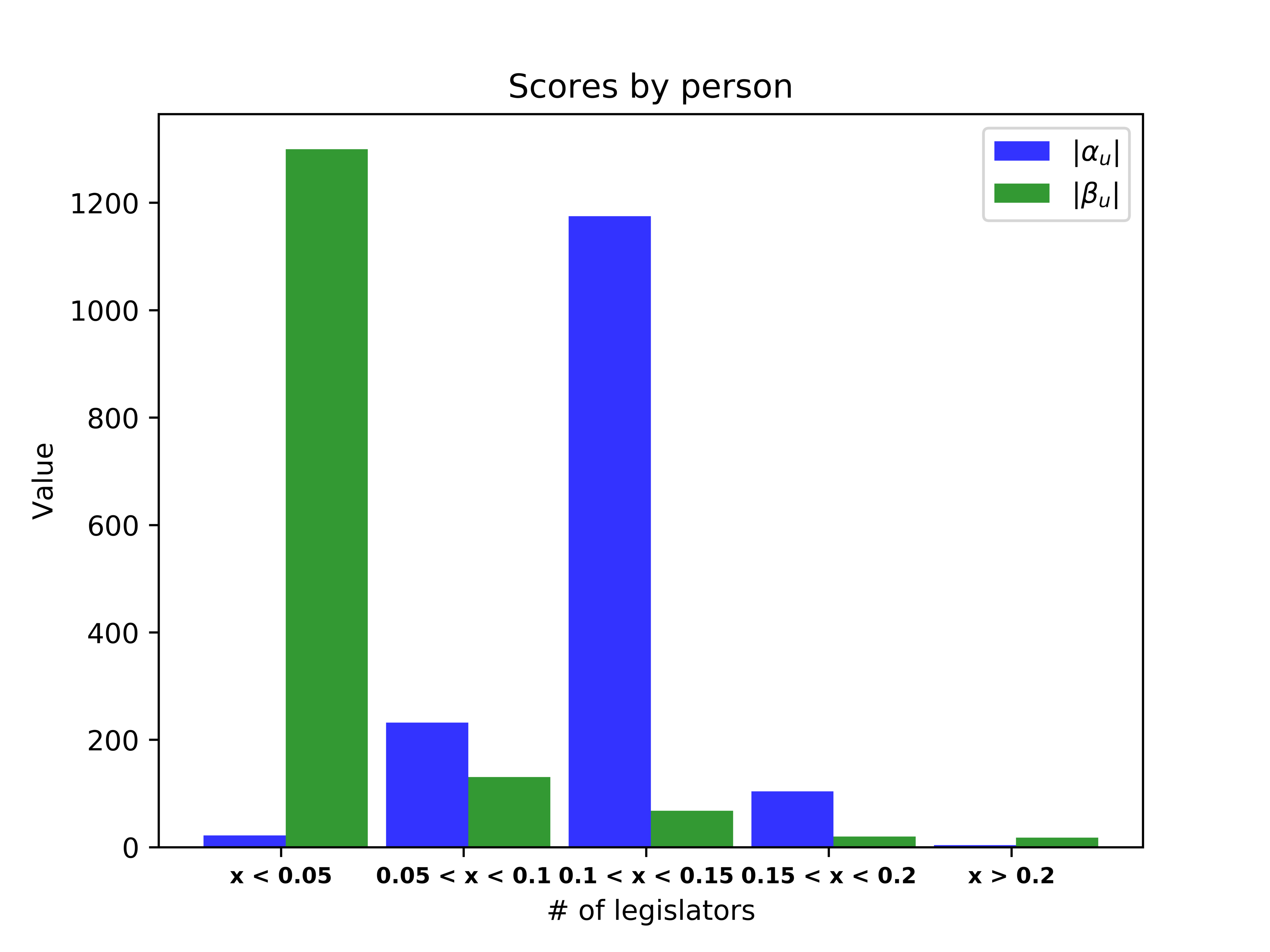}
	\caption{Graphical model representation of NIPEN-PGM}
	\label{fig:alpha_beta_individual}
\end{figure}

\begin{figure}[h!]
	\centering
	\includegraphics[width=3.2in,height=2.5in]{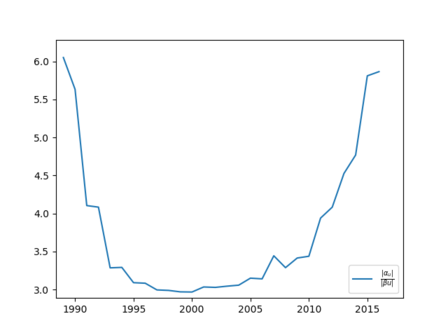}
	\caption{Graphical model representation of NIPEN-PGM}
	\label{fig:year_alpha_beta_ratio}
\end{figure}

According to the opinion of the reviewers who advised on the analysis of $\alpha_{u}$ and $\beta_{u}$, the distribution of $|\alpha_{u}|$ and $|\beta_{u}|$ is shown in Figure \ref{fig:alpha_beta_individual}, and the value of $|\alpha_{u}|/|\beta_{u}|$ over time is shown in Figure \ref{fig:year_alpha_beta_ratio}. According to Figure \ref{fig:alpha_beta_individual}, we can infer that the majority of legislators are voting focusing on contents rather than network effect. However, since the the number of element that $|\beta_{u}|>0.2$ is larger than the number of $|\alpha_{u}|>0.2$, we can infer that small number of legislators are highly dependent on network effect. In addition, Figure \ref{fig:year_alpha_beta_ratio} shows the change in a $|\alpha_{u}/\beta_{u}|$ over time, and we can infer that the influence of contents over networks has become more important in recent years.

%\clearpage
\selectfont
\bibliography{kw_aaai_ref}

\begin{thebibliography}{}

\bibitem[\protect\citeauthoryear{Abadi \bgroup et al\mbox.\egroup
  }{2016}]{abadi2016tensorflow}
Abadi, M.; Agarwal, A.; Barham, P.; Brevdo, E.; Chen, Z.; Citro, C.; Corrado,
  G.~S.; Davis, A.; Dean, J.; Devin, M.; and Others.
\newblock 2016.
\newblock {Tensorflow: Large-scale machine learning on heterogeneous
  distributed systems}.
\newblock {\em arXiv preprint arXiv:1603.04467}.

\bibitem[\protect\citeauthoryear{Bertero \bgroup et al\mbox.\egroup
  }{2016}]{berteroreal}
Bertero, D.; Siddique, F.~B.; Wu, C.-S.; Wan, Y.; Chan, R. H.~Y.; and Fung, P.
\newblock 2016.
\newblock {Real-Time Speech Emotion and Sentiment Recognition for Interactive
  Dialogue Systems}.
\newblock {\em ACL}.

\bibitem[\protect\citeauthoryear{Blei, Ng, and Jordan}{2003}]{Blei2003}
Blei, D.~M.; Ng, A.~Y.; and Jordan, M.~I.
\newblock 2003.
\newblock {Latent Dirichlet Allocation}.
\newblock {\em The Journal of Machine Learning Research} 3:993--1022.

\bibitem[\protect\citeauthoryear{Chaney, Blei, and
  Eliassi-Rad}{2015}]{Chaney2015}
Chaney, A.~J.; Blei, D.~M.; and Eliassi-Rad, T.
\newblock 2015.
\newblock {A probabilistic model for using social networks in personalized item
  recommendation}.
\newblock {\em Proceedings of the 9th ACM Conference on Recommender Systems}
  43--50.

\bibitem[\protect\citeauthoryear{Chen \bgroup et al\mbox.\egroup
  }{2012}]{Chen2012}
Chen, M.; Xu, Z.; Weinberger, K.; and Sha, F.
\newblock 2012.
\newblock {Marginalized Denoising Autoencoders for Domain Adaptation}.
\newblock {\em Proceedings of the 29th International Conference on Machine
  Learning (ICML)}  767----774.

\bibitem[\protect\citeauthoryear{Chong \bgroup et al\mbox.\egroup
  }{2017}]{chong2015predicting}
Chong, A. Y.~L.; Ch'ng, E.; Liu, M.~J.; and Li, B.
\newblock 2017.
\newblock {Predicting consumer product demands via Big Data: the roles of
  online promotional marketing and online reviews}.
\newblock {\em International Journal of Production Research} 55(17):5142--5156.

\bibitem[\protect\citeauthoryear{Clinton, Jackman, and
  Rivers}{2004}]{Clinton2004}
Clinton, J.~D.; Jackman, S.; and Rivers, D.
\newblock 2004.
\newblock {The Statistical Analysis of Roll Call Data}.
\newblock {\em The American Political Science Review} 98(2):355--370.

\bibitem[\protect\citeauthoryear{Cohen and Malloy}{2014}]{cohen2014friends}
Cohen, L., and Malloy, C.~J.
\newblock 2014.
\newblock {Friends in high places}.
\newblock {\em American Economic Journal: Economic Policy} 6(3):63--91.

\bibitem[\protect\citeauthoryear{Faust and Skvoretz}{2002}]{Faust2002}
Faust, K., and Skvoretz, J.
\newblock 2002.
\newblock {Comparing Networks Across Space and Time, Size and Species}.
\newblock {\em Networks} 32(2002):267--299.

\bibitem[\protect\citeauthoryear{Fowler}{2006}]{Fowler2006}
Fowler, J.~H.
\newblock 2006.
\newblock {Connecting the congress: A study of cosponsorship networks}.
\newblock {\em Political Analysis} 14(4):456--487.

\bibitem[\protect\citeauthoryear{Gerrish and Blei}{2012}]{Gerrish2012}
Gerrish, S., and Blei, D.~M.
\newblock 2012.
\newblock {How they vote: Issue-adjusted models of legislative behavior}.
\newblock {\em Advances in Neural Information Processing Systems}
  25(1):2762--2770.

\bibitem[\protect\citeauthoryear{Gu \bgroup et al\mbox.\egroup }{2014}]{Gu2014}
Gu, Y.; Sun, Y.; Jiang, N.; Wang, B.; and Chen, T.
\newblock 2014.
\newblock {Topic-factorized ideal point estimation model for legislative voting
  network}.
\newblock {\em Proceedings of the 20th ACM SIGKDD international conference on
  Knowledge discovery and data mining. ACM, 2014.}  183--192.

\bibitem[\protect\citeauthoryear{Guo, Zhang, and Yorke-Smith}{2015}]{Guo2015}
Guo, G.; Zhang, J.; and Yorke-Smith, N.
\newblock 2015.
\newblock {TrustSVD : Collaborative Filtering with Both the Explicit and
  Implicit Influence of User Trust and of Item Ratings}.
\newblock {\em Proceedings of the Twenty-ninth AAAI Conference on Artificial
  Intelligence (AAAI)}  123--129.

\bibitem[\protect\citeauthoryear{Heckman and {Snyder Jr}}{1996}]{Heckman1996}
Heckman, J.~J., and {Snyder Jr}, J.~M.
\newblock 1996.
\newblock {Linear probability models of the demand for attributes with an
  empirical application to estimating the preferences of legislators}.
\newblock {\em National bureau of economic research} 28(0).

\bibitem[\protect\citeauthoryear{Hofmann}{1999}]{Hofmann1999}
Hofmann, T.
\newblock 1999.
\newblock {Probabilistic latent semantic indexing}.
\newblock {\em Proceedings of the 22nd annual international ACM SIGIR
  conference on Research and development in information retrieval}  50--57.

\bibitem[\protect\citeauthoryear{Islam \bgroup et al\mbox.\egroup
  }{2016}]{Islam2016}
Islam, M.~R.; Hossain, K.~T.; Krishnan, S.; and Ramakrishnan, N.
\newblock 2016.
\newblock {Inferring Multi-dimensional Ideal Points for US Supreme Court
  Justices}.
\newblock {\em Proceedings of the 30th Conference on Artificial Intelligence
  (AAAI)}  4--12.

\bibitem[\protect\citeauthoryear{Kingma and Welling}{2014}]{Kingma2013}
Kingma, D.~P., and Welling, M.
\newblock 2014.
\newblock {Auto-encoding variational bayes}.
\newblock {\em In Proceedings of the International Conference on Learning
  Representations (ICLR).}

\bibitem[\protect\citeauthoryear{Koren, Bell, and Volinsky}{2009}]{Koren2009}
Koren, Y.; Bell, R.; and Volinsky, C.
\newblock 2009.
\newblock {Matrix Factorization Techniques for Recommender Systems}.
\newblock {\em Computer} 42(8):42--49.

\bibitem[\protect\citeauthoryear{Lafferty and
  Blei}{2006}]{lafferty2006correlated}
Lafferty, J.~D., and Blei, D.~M.
\newblock 2006.
\newblock {Correlated topic models}.
\newblock In {\em Advances in neural information processing systems},
  147--154.

\bibitem[\protect\citeauthoryear{Lee \bgroup et al\mbox.\egroup
  }{2013}]{Lee2013}
Lee, J.; Kim, S.; Lebanon, G.; and Singer, Y.
\newblock 2013.
\newblock {Local Low-Rank Matrix Approximation}.
\newblock {\em ICML} 28.

\bibitem[\protect\citeauthoryear{Li, Kawale, and Fu}{2015}]{Fu2015}
Li, S.; Kawale, J.; and Fu, Y.
\newblock 2015.
\newblock {Deep collaborative filtering via marginalized denoising
  auto-encoder}.
\newblock In {\em Proceedings of the 24th ACM International on Conference on
  Information and Knowledge Management},  811--820.
\newblock ACM.

\bibitem[\protect\citeauthoryear{Poole and Rosenthal}{1985}]{Poole1985}
Poole, K.~T., and Rosenthal, H.
\newblock 1985.
\newblock {A spatial model for legislative roll call analysis}.
\newblock {\em American Journal of Political Science}  357--384.

\bibitem[\protect\citeauthoryear{Ramage \bgroup et al\mbox.\egroup
  }{2009}]{Ramage2009}
Ramage, D.; Hall, D.; Nallapati, R.; and Manning, C.~D.
\newblock 2009.
\newblock {Labeled LDA: A supervised topic model for credit attribution in
  multi-labeled corpora}.
\newblock {\em Proceedings of the 2009 Conference on Empirical Methods in
  Natural Language Processing} 1(August):248--256.

\bibitem[\protect\citeauthoryear{Sedhain \bgroup et al\mbox.\egroup
  }{2015}]{Sedhain2015}
Sedhain, S.; Menon, A.~K.; Sanner, S.; and Xie, L.
\newblock 2015.
\newblock {AutoRec : Autoencoders Meet Collaborative Filtering}.
\newblock {\em Proceedings of the 24th International Conference on World Wide
  Web (WWW)}  111--112.

\bibitem[\protect\citeauthoryear{Shah, Rao, and Ding}{2017}]{shah2017matrix}
Shah, V.; Rao, N.; and Ding, W.
\newblock 2017.
\newblock {Matrix Factorization with Side and Higher Order Information}.
\newblock {\em arXiv preprint arXiv:1705.02047}.

\bibitem[\protect\citeauthoryear{Socher \bgroup et al\mbox.\egroup
  }{2013}]{socher2013reasoning}
Socher, R.; Chen, D.; Manning, C.~D.; and Ng, A.
\newblock 2013.
\newblock {Reasoning with neural tensor networks for knowledge base
  completion}.
\newblock In {\em Advances in neural information processing systems},
  926--934.

\bibitem[\protect\citeauthoryear{Vincent \bgroup et al\mbox.\egroup
  }{2010}]{VincentPASCALVINCENT2010}
Vincent, P.; Larochelle, H.; Lajoie, I.; Bengio, Y.; and Manzagol, P.-A.
\newblock 2010.
\newblock {Stacked Denoising Autoencoders: Learning Useful Representations in a
  Deep Network with a Local Denoising Criterion}.
\newblock {\em Journal of Machine Learning Research} 11:3371--3408.

\bibitem[\protect\citeauthoryear{Wang and Blei}{2011}]{ChongWang2011}
Wang, C., and Blei, D.~M.
\newblock 2011.
\newblock {Collaborative topic modeling for recommending scientific articles}.
\newblock In {\em Proceedings of the 17th ACM SIGKDD international conference
  on Knowledge discovery and data mining},  448--456.
\newblock ACM.

\bibitem[\protect\citeauthoryear{Wang, Shi, and Yeung}{2017}]{Wang2017}
Wang, H.; Shi, X.; and Yeung, D.-y.
\newblock 2017.
\newblock {Relational Deep Learning : A Deep Latent Variable Model for Link
  Prediction}.
\newblock {\em AAAI}.

\bibitem[\protect\citeauthoryear{Wang, Wang, and Yeung}{2015}]{Wang2015}
Wang, H.; Wang, N.; and Yeung, D.-Y.
\newblock 2015.
\newblock {Collaborative Deep Learning for Recommender Systems}.
\newblock {\em KDD}  1235--1244.

\bibitem[\protect\citeauthoryear{Wu \bgroup et al\mbox.\egroup }{2016}]{Wu2016}
Wu, Y.; DuBois, C.; Zheng, A.~X.; and Ester, M.
\newblock 2016.
\newblock {Collaborative Denoising Auto-Encoders for Top-N Recommender
  Systems}.
\newblock {\em Proceedings of the Ninth ACM International Conference on Web
  Search and Data Mining - WSDM '16}  153--162.

\bibitem[\protect\citeauthoryear{Ying \bgroup et al\mbox.\egroup
  }{2016}]{Ying2016}
Ying, H.; Chen, L.; Xiong, Y.; and Wu, J.
\newblock 2016.
\newblock {Collaborative deep ranking: A hybrid pair-wise recommendation
  algorithm with implicit feedback}.
\newblock {\em Pacific-Asia Conference on Knowledge Discovery and Data Mining}
  9652 LNAI:555--567.

\end{thebibliography}
\bibliographystyle{aaai}
\end{document}